\documentclass[twocolumn, tighten, numberedappendix, twocolappendix]{aastex63}

\usepackage[utf8]{inputenc}
\usepackage{graphicx}      
\usepackage{amsmath,amstext}
\usepackage{ulem}
\usepackage{epstopdf}

\providecommand{\sorthelp}[1]{}

\usepackage{xcolor}

\usepackage[encapsulated]{CJK}
\usepackage{ucs}
\newcommand{\cntext}[1]{\begin{CJK}{UTF8}{gbsn}#1\end{CJK}}

\usepackage{devanagari}

\graphicspath{{./}{figures/}}

\newcommand{\nTotalTimestreams}{15,342 }
\newcommand{\nTotalDetdays}{2,557 }

\shorttitle{CLASS 40~GHz VPM Demodulation Stability}
\shortauthors{Harrington et al.}

\newcommand{\jhu}{Department of Physics and Astronomy, Johns Hopkins University, 
3701 San Martin Drive, Baltimore, MD 21218, USA}
\newcommand{\nasa}{Goddard Space Flight Center, 8800 Greenbelt Road, Greenbelt, MD 20771, USA}

\begin{document}
\title{Two Year Cosmology Large Angular Scale Surveyor (CLASS) Observations: Long Timescale Stability Achieved with a Front-End Variable-delay Polarization Modulator at 40~GHz}

\correspondingauthor{Kathleen Harrington}
\email{katieharrington@uchicago.edu}

\author[0000-0003-1248-9563]{Kathleen Harrington}
\affil{Department of Astronomy and Astrophysics, University of Chicago, 5640 South Ellis Avenue, Chicago, IL 60637, USA}
\affil{Department of Physics, University of Michigan, 450 Church St, Ann Arbor, MI, 48109, USA}
\affil{\jhu}

\author[0000-0003-3853-8757]{Rahul Datta}
\affil{\jhu}

\author[0000-0003-2838-1880]{Keisuke Osumi}
\affil{\jhu}

\author[0000-0001-7941-9602]{Aamir Ali}
\affiliation{Department of Physics, University Of California, Berkeley, CA 94720, USA}
\affil{\jhu}

\author[0000-0002-8412-630X]{John~W. Appel}
\affil{\jhu}

\author[0000-0001-8839-7206]{Charles~L. Bennett}
\affil{\jhu}

\author{Michael~K. Brewer}
\affil{\jhu}

\author[0000-0001-8468-9391]{Ricardo Bustos}
\affil{Departamento de Ingenier\'ia El\'ectrica, Universidad Cat\'olica de la Sant\'isima Concepci\'on, Alonso de Ribera 2850, Concepci\'on, Chile}

\author[0000-0003-1127-0965]{Manwei Chan}
\affiliation{\jhu}

\author[0000-0003-0016-0533]{David T.~Chuss}
\affiliation{Department of Physics, Villanova University, 800 Lancaster Avenue, Villanova, PA 19085, USA}

\author{Joseph Cleary}
\affil{\jhu}

\author[0000-0002-0552-3754]{Jullianna Denes~Couto}
\affil{\jhu}

\author[0000-0002-1708-5464]{Sumit Dahal ({\dn \7{s}Emt dAhAl})}
\affil{\nasa}
\affil{\jhu}

\author{Rolando D\"unner}
\affil{Instituto de Astrof\'isica and Centro de Astro-Ingenier\'ia, Facultad de F\'isica, Pontificia Universidad Cat\'olica de Chile, Av. Vicu\~na Mackenna 4860, 7820436 Macul, Santiago, Chile}

\author[0000-0001-6976-180X]{Joseph R.~Eimer}
\affiliation{\jhu}

\author[0000-0002-4782-3851]{Thomas~Essinger-Hileman}
\affil{\nasa}

\author[0000-0002-2781-9302]{Johannes Hubmayr}
\affil{Quantum Sensors Group, National Institute of Standards and Technology, 325 Broadway, Boulder, CO 80305, USA}

\author{Francisco Raul~Espinoza~Inostroza}
\affil{Departamento de Ingenier\'ia El\'ectrica, Universidad Cat\'olica de la Sant\'isima Concepci\'on, Alonso de Ribera 2850, Concepci\'on, Chile}

\author[0000-0001-7466-0317]{Jeffrey Iuliano}
\affil{\jhu}

\author{John Karakla}
\affiliation{\jhu}

\author[0000-0002-4820-1122]{Yunyang Li (\cntext{李云炀}\!\!)}
\affil{\jhu} 

\author[0000-0003-4496-6520]{Tobias A.~Marriage}
\affil{\jhu}

\author{Nathan J.~Miller}
\affil{\jhu}
\affil{\nasa}

\author[0000-0002-5247-2523]{Carolina N\'u\~{n}ez}
\affil{\jhu}

\author[0000-0002-0024-2662]{Ivan L.~Padilla}
\affil{\jhu}

\author[0000-0002-8224-859X]{Lucas Parker}
\affiliation{Space and Remote Sensing, MS B244, Los Alamos National Laboratory, Los Alamos, NM 87544, USA}
\affiliation{\jhu}

\author[0000-0002-4436-4215]{Matthew A. Petroff}
\affil{\jhu}

\author[0000-0001-5167-7159 ]{Bastian Pradenas~M\'{a}rquez}
\affil{\jhu}
\affil{Departamento de F\'isica, FCFM, Universidad de Chile, Blanco Encalada 2008, Santiago, Chile}

\author[0000-0001-5704-271X]{Rodrigo Reeves}
\affil{CePIA, Departamento de Astronomia, Universidad de Concepcion, Concepcion, Chile}

\author[0000-0002-2061-0063]{Pedro Flux\'a Rojas}
\affiliation{Instituto de Astrof\'isica, Facultad de F\'isica, Pontificia Universidad Cat\'olica de Chile, Avenida Vicu\~na Mackenna 4860, 7820436, Chile}
\affiliation{Instituto de Astrof\'isica and Centro de Astro-Ingenier\'ia, Facultad de F\'isica, Pontificia Universidad Cat\'olica de Chile, Av. Vicu\~na Mackenna 4860, 7820436 Macul, Santiago, Chile}

\author[0000-0003-4189-0700]{Karwan Rostem}
\affil{\nasa}

\author[0000-0003-3487-2811]{Deniz Augusto~Nunes~Valle}
\affil{\jhu}

\author[0000-0002-5437-6121]{Duncan J.~Watts}
\affil{Institute of Theoretical Astrophysics, University of Oslo, Blindern, Oslo, Norway}
\affil{\jhu}

\author[0000-0003-3017-3474]{Janet L.~Weiland}
\affil{\jhu}

\author[0000-0002-7567-4451]{Edward J.~Wollack}
\affil{\nasa}

\author[0000-0001-5112-2567]{Zhilei Xu (\cntext{徐智磊}\!\!)}
\affiliation{MIT Kavli Institute, Massachusetts Institute of Technology, Cambridge, Massachusetts 02139, USA}
\affiliation{Department of Physics and Astronomy, University of Pennsylvania, 209 South 33rd Street, Philadelphia, PA 19104, USA}
\affil{\jhu}

\collaboration{36}{CLASS Collaboration}

\begin{abstract}
The Cosmology Large Angular Scale Surveyor (CLASS) is a four-telescope array observing the largest angular scales ($2 \lesssim \ell \lesssim 200$) of the cosmic microwave background (CMB) polarization. These scales encode information about reionization and inflation during the early universe. The instrument stability necessary to observe these angular scales from the ground is achieved through the use of a variable-delay polarization modulator (VPM) as the first optical element in each of the CLASS telescopes. Here we develop a demodulation scheme used to extract the polarization timestreams from the CLASS data and apply this method to selected data from the first two years of observations by the 40~GHz CLASS telescope. These timestreams are used to measure the $1/f$ noise and temperature-to-polarization ($T\rightarrow P$) leakage present in the CLASS data. We find a median knee frequency for the pair-differenced demodulated linear polarization of 15.12~mHz and a $T\rightarrow P$ leakage of $<3.8\times10^{-4}$ (95\% confidence) across the focal plane. We examine the sources of $1/f$ noise present in the data and find the component of $1/f$ due to atmospheric precipitable water vapor (PWV) has an amplitude of $203 \pm 12~\mathrm{\mu K_{RJ}\sqrt{s}}$ for 1~mm of PWV when evaluated at 10~mHz; accounting for $\sim32\%$ of the $1/f$ noise in the central pixels of the focal plane. The low level of $T\rightarrow P$ leakage and $1/f$ noise achieved through the use of a front-end polarization modulator enables the observation of the largest scales of the CMB polarization from the ground by the CLASS telescopes.
\end{abstract}

\keywords{\href{http://astrothesaurus.org/uat/799}{Astronomical instrumentation (799)}; \href{http://astrothesaurus.org/uat/1127}{Polarimeters (1127)}; \href{http://astrothesaurus.org/uat/322}{Cosmic microwave background radiation (322)}; \href{http://astrothesaurus.org/uat/435}{Early Universe (435)}; \href{http://astrothesaurus.org/uat/1146}{Observational Cosmology (1146)};  }

\section{Introduction}

Observations of the cosmic microwave background (CMB) have revolutionized our understanding of the universe and established $\mathrm{\Lambda}$CDM as the standard model of cosmology \citep{wmap_yr1_bennett,wmap_yr1_spergel,bennett:2013}. Much of the cosmological information in the CMB is encoded in its angular power spectrum, which describes the amplitude of the CMB anisotropy as a function of angular scale or multipole moment, $\ell$ \citep{peeb70,doro78}. While the large scale CMB temperature anisotropy has been measured to below the cosmic variance limit by the space satellites \textit{WMAP} \citep{Hinshaw2013} and \textit{Planck} \citep{planck2016-l01}, measurements of the polarization anisotropy on the same scales have yet to reach this limit. 

The largest angular scales of the CMB polarization contain a wealth of information from the early universe. The amplitude of the E-modes on largest scales ($\ell \lesssim 20$) depends on the optical depth to reionization, $\tau$ \citep{Page2007,planck2016-l06}. The shape of the largest scale E-modes depends on the specific reionization history, such that the free electron fraction over cosmic time can be constrained using high signal-to-noise large-scale E-mode measurements \citep{hu, heinrich, watts2020}. Additionally, constraining specific inflationary models using large-scale B-mode measurements requires excellent knowledge of the reionization history \citep{obied}. Observing the large scale E-modes requires measuring angular correlations on large fractions of sky to better than $\sim200$~nK. 

The large angular scale B-modes are expected to have an amplitude proportional to the tensor-to-scalar ratio $r$. A non-zero tensor-to-scalar ratio would indicate the presence of super-horizon gravitational waves predicted to have been produced during an inflationary expansion phase in the early universe \citep{kami97,Sel_Zal1997}. The strongest constraint to date on the tensor-to-scalar ratio from CMB B-mode and temperature measurements is $r<0.06$ at 95\% confidence \citep{bicep2_2018_2015}. For $r=0.01$, detecting the large angular scale B-modes requires an angular correlation measurement at the level of $\sim30$~nK. 

Observations of the largest angular scales are limited in part by scan strategy and the stability of the instrument. Mapping the lowest multipole moments requires a scan strategy that covers large angles ($\theta\sim2\pi/\ell$) while the telescope noise remains sufficiently stationary. Noise on long timescales generally has a $1/f$ spectrum with a knee frequency, where the $1/f$ noise equals the white noise, that parameterizes the rate of change of the instrument noise. Temperature drifts in the atmosphere and instrument are the dominant $1/f$ noise sources for intensity observations. These fluctuations are converted into polarization due to temperature to polarization ($T\rightarrow P$) leakage resulting from non-idealities in the instrument. While $T \rightarrow P$ leakage in polarization optimized instruments is usually small, $\lesssim1\%$ \citep{essinger2016}, it can still be a dominant effect when integrating down to the sensitivity level necessary to observe the polarization of the CMB. Additionally, non-optical sources of $1/f$ from the cryogenics, detectors, and readout system primarily affect the intensity signal but can leak into polarization depending on the exact source and analysis method.

Mapping the largest scales on the sky requires a method of minimizing the effects of $T\rightarrow P$ leakage and $1/f$ more broadly. Space satellites are able to scan the sky very quickly, with $f_\mathrm{scan} \sim 20$~mHz ($\sim 1$~RPM), and are unaffected by atmosphere. When the scan frequency is below or close to the $1/f$ knee frequency, destriping methods such as MADAM \citep{madam2010} can be used to filter out the long timescale fluctuations while retaining most of the sky signal. If the scan frequency is significantly slower than the knee frequency, the signal from large angular scale CMB fluctuations will be overwhelmed by the $1/f$ noise. In that regime, filtering out the noise without also filtering out the sky signal is not feasible \citep{poletti2016}.

Ground-based CMB observations are usually executed in a regime such that the scan frequency with respect to the largest angular scales is much lower than the $1/f$ knee frequency. Diurnal temperature variations and $1/f$ noise from the atmosphere make the ground observation environment inherently less stable than space observations. For these and other reasons, ground-based CMB surveys have usually restricted their observations to smaller fractions of sky and/or filtered out the longer timescale modes needed to map the largest angular scales. 

Several ground-based instruments have been experimenting with polarization modulation to increase long-time-scale stability and reduce $T \rightarrow P$ leakage. This has been particularly important in the Chilean Atacama Desert, where large sky fractions are accessible but the atmosphere is less stable than at the South Pole. QUIET \citep{QUIET_results2011} used phase modulation with coherent detection techniques to reduce the $1/f$ noise in their timestreams. ABS \citep{abs_final}, \textsc{Polarbear} \citep{adac20}, and ACTPol \citep{loui17} have all implemented or tested continuously rotating half waveplate (HWP) modulators. The HWP modulators have generally increased instrument stability and reduced $T \rightarrow P$ leakage \citep{abs_hwp, Takakura2017}; however, the scan strategies implemented by these experiments focused only on mapping angular fluctuations on degree scales and smaller. 

The Cosmology Large Angular Scale Surveyor (CLASS) \citep{essi14, harr16} is a unique project targeting the largest angular scales ($1^\circ \lesssim \theta \lesssim 90^\circ$) of the CMB polarization from the ground. Sited in the Chilean Atacama desert \citep{parque_atacama}, CLASS observes in four frequency bands: 40, 90, 150, and 220~GHz. This set of frequency bands spans the Galactic foreground minimum and operate in spectral windows of high atmospheric transmission. During the first two years considered in this paper, CLASS observed continuously, executing constant $45^\circ$ elevation scans through $720^\circ$ in azimuth at $1^\circ\,\mathrm s^{-1}$. With this scan strategy and site location, CLASS observed nearly 75\% of the sky every day. Daily boresight rotations, in $15^\circ$ increments through $\pm45^\circ$, are used to rotate the polarization sensitivity of the instrument \citep{essi14}.

\begin{figure}
\includegraphics[width=0.9\columnwidth]{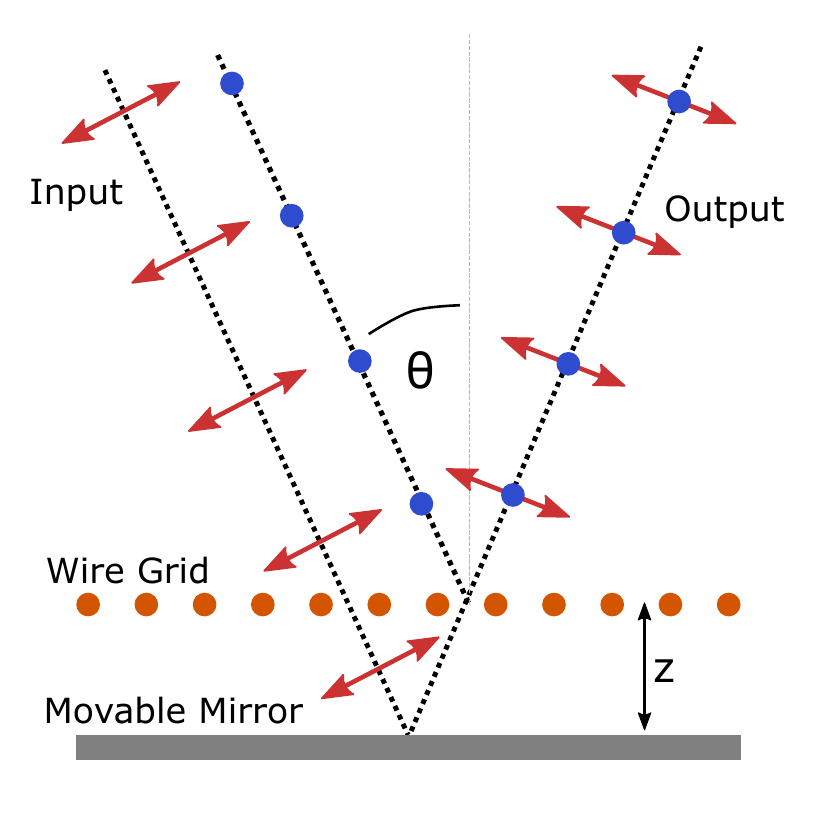}
\caption{\label{fig:vpm_sketch} A schematic diagram of a variable-delay polarization modulator as implemented for the CLASS telescopes. A linearly polarizing wire grid is mounted in front of a movable mirror. This creates a phase delay between polarization states parallel and perpendicular to the direction of the grid wires. Changing the distance between the wire grid and the mirror, $z$, results in a modulation between one of the linear Stokes polarizations and circular polarization.}
\end{figure}

The unique aspect of the CLASS telescope is the use of a variable-delay polarization modulator (VPM) as the first optical element seen by radiation coming from the sky \citep{chus12, mill15, harr18}. Sketched in Figure \ref{fig:vpm_sketch}, a VPM consists of a linearly polarizing wire grid positioned in front of and parallel to a movable mirror. The distance between the grid and the mirror creates a phase delay between polarization states parallel and perpendicular to the wires. Moving the mirror position with respect to the wire grid changes this phase delay and modulates one linear polarization and circular polarization. For CLASS, this modulation is rapid, with the mirror moving at 10~Hz. This encodes the sky polarization signals at frequencies well above the $1/f$ knee frequencies of the detectors and atmosphere. In addition, having the VPM as the first optical element means the polarization modulation occurs before the effects of instrument polarization from the later optical elements. 

Here we discuss the demodulation process used to extract the polarization timestreams from the CLASS data, present results on the achieved instrument stability and $T\rightarrow P$ leakage, and discuss the sources of $1/f$ noise in the demodulated data. This paper is part of a series of papers on the first 2~years (Sept~2016 to March~2018, ``Era 1'' hereafter) of observations by the 40~GHz CLASS telescope. Other papers in this series cover the calibration \citep{appe19}, beams \citep{xu20}, pipeline (Osumi et al. In preparation), and results (Eimer et al. In preparation). In addition, this dataset has been used to put limits on the amplitude of the cosmological circular polarization \citep{padi20}, and detect atmospheric circular polarization at 40~GHz for the first time \citep{petr20}.

\section{\label{sec:demod}Demodulation}

An ideal VPM is one where all surfaces are lossless and the wavelengths of incident light are much longer than the wire diameters and pitch. In this regime the phase difference, $\delta$, between polarization states parallel and perpendicular to the wires depends on the frequency, $\nu$, the angle of incidence, $\theta$, and the grid-mirror distance, $z$, as

\begin{equation}
    \delta = \frac{4\pi \nu}{c} z \cos \theta.
\end{equation}

\noindent The phase delay between orthogonal linear polarization states creates a mixing between one of the linear Stokes parameters and circular polarization. In a $\mathrm{VPM}$ centered coordinate system, with Stokes $\pm Q$ defined parallel and perpendicular to the wire grid and plane of incidence perpendicular to the wire grid, the single-frequency Mueller matrix \citep{hecht2012optics} describing the relation between the input and output Stokes parameters is

\begin{equation}
M_\mathrm{VPM}(\nu,z) = \begin{bmatrix}
1 & 0 & 0 & 0 \\
0 & 1 & 0 & 0 \\
0 & 0 & \cos \delta(\nu,z)& -\sin \delta(\nu,z) \\
0 & 0 & \sin \delta(\nu,z) & \cos \delta(\nu,z) \\
\end{bmatrix}.
\end{equation}

\noindent The frequency-integrated, VPM-centered Mueller matrix can be calculated by integrating each element over the bandpass of the detectors and optics, $B(\nu)$. For example, the $UU$ element of the matrix is

\begin{equation}
(M_\mathrm{VPM})_{UU}(z) = \int d\nu \, B(\nu) \,  \cos \delta(\nu, z). 
\end{equation}


To calculate the modulation observed at a detector, the VPM-centered Mueller matrix is rotated into the detector-centered coordinate system that was described in Section 5 of \cite{xu20}. Critically, this rotation is different for every pixel in the focal plane. The VPM is near the entrance pupil of the CLASS optical design \citep{eime12}, meaning each detector illuminates the same area of the VPM, but each pixel has a different angle and plane of incidence on the VPM.\footnote{The different angles and planes of incidence also change the VPM centered Mueller matrix for each pixel.} From this coordinate system, the polarization of the detectors is well defined; all detectors are aligned along Stokes $\pm U$. A set of modulation functions, $S_{I}$, $S_{Q}$, $S_{U}$, and $S_{V}$, are calculated for each detector to describe the level at which each incoming Stokes parameter is modulated by the VPM. The power observed by each detector as a function of time, $t$, and grid-mirror distance, $z$, is

\begin{equation}
\begin{aligned}
    d(z,t) &= A_{z}(z)T_\mathrm{VPM}(t) 
    + S_{I}(z)I_\mathrm{in}(t)\\
    &+ S_{Q}(z)Q_\mathrm{in}(t)
    + S_{U}(z)U_\mathrm{in}(t)\\
    &+ S_{V}(z)V_\mathrm{in}(t).
\end{aligned}
\label{eq:data_vec}
\end{equation}

\noindent Here, the ``$\mathrm{in}$'' subscripts indicate the Stokes parameters of the light incident on the VPM in the detector coordinate system, and we have added the VPM temperature $T_\mathrm{VPM}$ and $A_z$ to indicate a VPM synchronous signal (VSS) that occurs due to emission from the wire grid and mirror \citep{mill15}.

For detectors with a non-zero azimuth offset in the CLASS focal plane, there is a rotation between Stokes ${Q}/{U}$ in the detector coordinate system and the linear polarization that is modulated by the VPM. For this reason, $S_{Q}$ and $S_{U}$ are highly covariant, and we define the linear polarization axes

\begin{equation}
\begin{aligned}
P_\mathrm{in}  &= -Q_\mathrm{in}\sin{2\phi_\mathrm{P}} + U_\mathrm{in}\cos{2\phi_\mathrm{P}} \\
X_\mathrm{in}  &= Q_\mathrm{in}\cos{2\phi_\mathrm{P}} + U_\mathrm{in}\sin{2\phi_\mathrm{P}}, \\
\end{aligned}
\label{eg:lin_axes}
\end{equation}

\noindent where $\phi_{P}$ is the axis of linear polarization that is modulated for each given detector. For the CLASS 40~GHz design, $|\phi_{P}| < 3^\circ$ across the focal plane. Implementing this rotation orthogonalizes $S_{Q}$ and $S_{U}$ into $S_{P}$ and $S_{X}$, where $S_{P}$ is the modulation of linear polarization along $\phi_{P}$ and $S_{X}$ is the sensitivity to the un-modulated polarization axis.

For an ideal VPM with zero instrument emissivity, unpolarized light and the $X$ linear polarization are unmodulated, meaning $A_z=0$, $S_{I}=1$, and $S_{X}$ is constant. For real VPMs, the emissivity of the VPM, $\epsilon_\mathrm{VPM}$, changes as a function of grid-mirror distance ($\Delta\epsilon_\mathrm{VPM} \sim 5\times10^{-4}$), due to the differential emissivity between the wire grid and the mirror. This causes aspects of $S_I$, $S_X$, and $A_z$ to change with grid-mirror distance; these combine to make up the VSS. The amplitude of the VSS is nearly constant, but it does depend on the temperature of the VPM, which drifts slowly over time. The portions of $S_{I}$ and $S_{X}$ that change with grid-mirror distance are expected to have identical shapes. Since any possible polarized atmospheric signal should be orders of magnitude below the intensity signal, we will ignore $S_{X}$ in the rest of this analysis.

\begin{figure*}
\includegraphics[width=\textwidth]{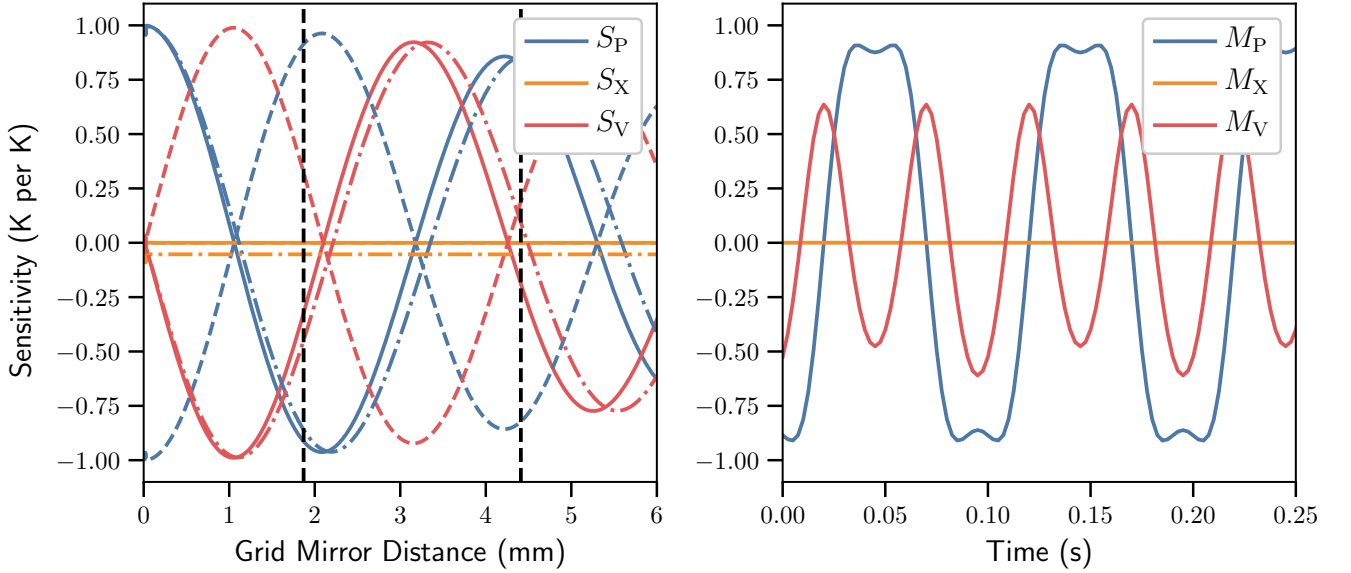}
\caption{\label{fig:mod_funct} The left plot shows three example sets of modulation functions for a $+45^\circ$ detector near the center of the focal plane (solid lines), the $-45^\circ$ detector in the same pixel (dashed lines), and a $+45^\circ$ detector near the top left edge of the focal plane (dot dashed lines). The black dashed lines denote the VPM throw for the 40~GHz telescope, the distances traversed by the VPM mirror during operation. The blue lines are $S_{P}$, the sensitivity to the modulated linear polarization as a function of VPM grid-mirror distance. The two orthogonal detectors in the same focal plane have modulation functions that have opposite signs. The pixel at the top of the focal plane, which has a larger angle of incidence on the VPM, has peaks slightly offset from the central pixels due to the larger path length through the VPM. The orange lines show $S_{X}$, which is the sensitivity to the unmodulated linear polarization. This is approximately zero for detectors with no azimuthal offset in the focal plane but is non-zero for detectors with an azimuthal offset due to the projection geometry onto the VPM. The red lines show $S_{V}$, which is out of phase with $S_{P}$ but has the same characteristics; orthogonal detectors have opposite modulation functions, and detectors at different elevation offsets in the focal plane have different peak locations. The right plot shows the modulation timestreams, $M$, (Equation \ref{eq:mod_time}) for the center $+45^\circ$ detector shown also in solid lines in the left plot. These timestreams use the 40~GHz VPM throw and modulation frequency (10~Hz) to convert the Stokes sensitivity as a function of grid-mirror distance to time.}
\end{figure*}

More complete derivations for these modulation functions are discussed in \cite{Harrington_Thesis_2018}. Figure \ref{fig:mod_funct} shows the modulation functions for three detectors, two orthogonal detectors near the center and one detector on the top edge of the focal plane. The modulation functions for the different detectors illustrate the effects of detector angle and incidence on the VPM. Orthogonal detectors in the same pixel have modulation functions with opposite signs, and detectors with different elevation and azimuth offsets have slightly shifted modulation functions due to the projection geometry onto the VPM and the different path lengths through the VPM.

The advantage of a continuous polarization modulator is that it moves the polarization signal to a frequency band that is above the $1/f$ knee of the system. To remove a large fraction of the correlated noise in the detector timestreams and to prevent the large, fluctuating intensity signal from numerically impacting the demodulation process, the first step in demodulation is to prefilter the data with a high pass or bandpass filter that preserves sufficient bandwidth around the modulation frequency but removes the long timescale drifts in the data. In practice, the exact shape of this filter does not impact the demodulated signals as long as the cutoff is above the $1/f$ knee frequency and the modulated signals are preserved.

Next, we define the modulation timestreams to be the VPM modulation functions over time. These are shown in the right plot of Figure \ref{fig:mod_funct} and are calculated using the array of VPM mirror positions that are synchronously sampled with the detector data. These modulation timestreams are prefiltered identically to the detector data,

\begin{equation}
    M_{i}(t) = \mathcal{F} \left\{ S_{i}\left[z(t) \right]\right\}.
\label{eq:mod_time}
\end{equation}

\noindent Here, $z(t)$ is the VPM grid-mirror distance as a function of time, $i$ denotes one of the different modulation functions described above, and $\mathcal{F}$ is the prefiltering applied. As discussed in \cite{harr18}, CLASS VPMs implement a sinusoidal mirror motion where the range of grid-mirror distances traversed by the VPM is set to maximize the time spent observing the linear polarization. Choosing this setting, combined with the effects of decoherence across the detector bandpass and the necessary pre-filtering of the timestream data, results in modulation timestreams that contain many harmonics of the modulation frequency, $f_\mathrm{mod}$. Critically, $M_{P}$ and $M_{V}$ are not orthogonal functions.

Demodulation of detector timestreams in the context of partially covariant modulation functions is accomplished through an explicit solution to a least-squares fit, 
\begin{equation}
\begin{bmatrix} P \\ V \end{bmatrix} = (\mathbf{M}^T \mathbf{N}^{-1} \mathbf{M})^{-1} \mathbf{M}^T \mathbf{N}^{-1}  \mathcal{F}\vec{d}    
\label{eq:least_sqr}
\end{equation}
\noindent where $\vec{d}$ is the detector timestream (Equation~\ref{eq:data_vec}) and $P$ is the linear polarization modulated by the VPM (Equation~\ref{eg:lin_axes}). In this fit, $\mathbf{N}$ is an estimate of the detector noise that is assumed to be white and uncorrelated due to the prefiltering of the data. The matrix $M$ has two columns containing the modulation timestreams for ${P}$ and ${V}$,

\begin{equation}
\mathbf{M} = 
\begin{bmatrix} 
M_{P}(t_0) & M_{V}(t_0) \\ 
M_{P}(t_1) & M_{V}(t_1) \\ 
\vdots & \vdots \\ 
M_{P}(t_N) & M_{V}(t_N) 
\end{bmatrix}.
\end{equation}

The contributions from $A_z$ and $S_{I}$ are removed because their amplitudes are several orders of magnitude smaller than $S_{P}$ and $S_{V}$, but their functional forms are similar in shape to $S_{P}$. Including these terms in the demodulation would make it difficult to reliably extract the polarization timestreams. Instead, these effects will be projected into the ${P}$ and ${V}$ timestreams, and any variations of their contributions will be constrained.

Using this least-squares fitting setup and assuming the detector noise is constant over the fit time period, the demodulated linear ($P$) and circular ($V$) timestreams are
\begin{equation}
\begin{aligned}
P = \mathcal{D}^{-1} &  \left\{ \textstyle{\sum} M_{V}^2 \; \textstyle{\sum} (M_{P}\,\mathcal{F}d) \right.  \\
 & \left. -\textstyle{\sum} (M_{P} M_{V}) \; \textstyle{\sum} (M_{V}\,\mathcal{F}d) \right\} 
\label{eq:p_demod}
\end{aligned}
\end{equation}

\noindent and

\begin{equation}
\begin{aligned}
V = \mathcal{D}^{-1} & \left\{ \textstyle{\sum} M_{P}^2 \; \textstyle{\sum} (M_{V}\,\mathcal{F}d) \right. \\
& \left. - \textstyle{\sum} (M_{P} M_{V}) \; \textstyle{\sum} (M_{P}\,\mathcal{F}d) \right\} 
\label{eq:v_demod}
\end{aligned}
\end{equation}

\noindent with determinant term

\begin{equation}
\mathcal{D} = \textstyle{\sum} M_{P}^2 \textstyle{\sum} M_{V}^2 - \left\{\textstyle{\sum} (M_{P} M_{V})\right\}^2.
\label{eq:determinant}
\end{equation}

\noindent The summations in Equations \ref{eq:p_demod}, \ref{eq:v_demod}, and \ref{eq:determinant} are for a finite period of time over which the polarization incident on the VPM is constant. The summations are replaced with a low-pass filter to produce a continuous demodulated timestream. This low-pass filter must have a cutoff frequency that is less than the modulation frequency, $f_\mathrm{mod}$, and greater than the rate of change of the incident sky signal, which depends on the beam width and scan speed of the instrument.

The inverse Fisher matrix, $\Sigma_{PV} = (\mathbf{M}^T \mathbf{N}^{-1} \mathbf{M})^{-1}$, is the covariance matrix between the two parameters. Tracking this quantity is necessary due to the covariance between the modulation functions and its dependence on the specific VPM positions traversed during the timestream.

A $3\times2$ rotation matrix, $R(\phi_{P})$ converts the demodulated timestreams and their associated covariance matrix into the $[Q, U, V]$ vectors, which are used in the mapping routine. While the VPM only modulates one of the linear Stokes parameters with circular polarization, boresight rotation and sky rotation enables full sampling of the Stokes parameters on the sky.

\section{\label{sec:methods}Data Selection and Processing}

A study was undertaken to constrain the long time scale noise properties of the CLASS data before and after demodulation by examining 2~hour long segments of data collected throughout Era~1 for the 40~GHz CLASS telescope. These segments were chosen during times where the Sun was at least $10^\circ$ below the horizon. This limited the impact of azimuth-synchronous signals and isolated time periods where Sun avoidance was unnecessary and, thus, the scan frequency was constant. They are also limited to when the precipitable water vapor (PWV) in the atmosphere, as measured by APEX\footnote{APEX PWV data was taken from their radiometer website,  \url{https://archive.eso.org/wdb/wdb/asm/meteo_apex/form}} or ALMA\footnote{ALMA PWV data, used for periods when the APEX radiometer was not running, was acquired via private correspondence.}, was less than 3~mm. In addition, to prevent a data glitch or jump repair from biasing the result, data segments were only used if less than $1\%$ of the timestreams were flagged by glitch and jump detection routines. 

Overall, \nTotalTimestreams two hour timestreams of detector pairs, representing \nTotalDetdays detector days of data, pass these criteria and are used in this analysis. This is significantly less than the total data acquired during the first era of 40~GHz observations. These selection criteria were chosen to enable a study of the intrinsic stability achieved by the VPM during this observational period, making it possible to quantify the benefits and limitations of the front-end VPM polarization modulator without confusion from other confounding systematics. The scientific analysis for this first era of 40~GHz observations will not be subject to the restriction of 2~hour sections of uninterrupted observing or the sun elevation limit, meaning the analyzed data set will be larger and a wider selection of potential systematic effects will be considered.

In addition to the aforementioned survey data, we also analyze a small set of stare data, taken when the telescope mount was stationary, from six 2-hour time periods during the first half of 2019. The stare data contains the $1/f$ noise associated with the VPM and the atmosphere, along a single line of sight, while avoiding any additional effects from mount motion or scan pattern. Many parts of the telescope structure were upgraded between Era~1 and the 2019 stare data, primarily due to the installation of the 90~GHz telescope. A second set of mirrors and receiver were installed on the mount, and the front-end forebaffle was expanded to serve both telescopes. In addition, the wire grid on the 40~GHz VPM was replaced with one with an improved copper finish on the wires \citep{harr18}, although this replacement does not appear to have significantly influenced the demodulated data compared to Era~1 data. Using data from the different observational configuration is necessary because no equivalent data are available from Era~1, where stare observations were shorter than 2~hours or during sub-par weather. 

\begin{figure*}
    \centering
    \begin{tabular}{c}
         \includegraphics[width=0.95\textwidth]{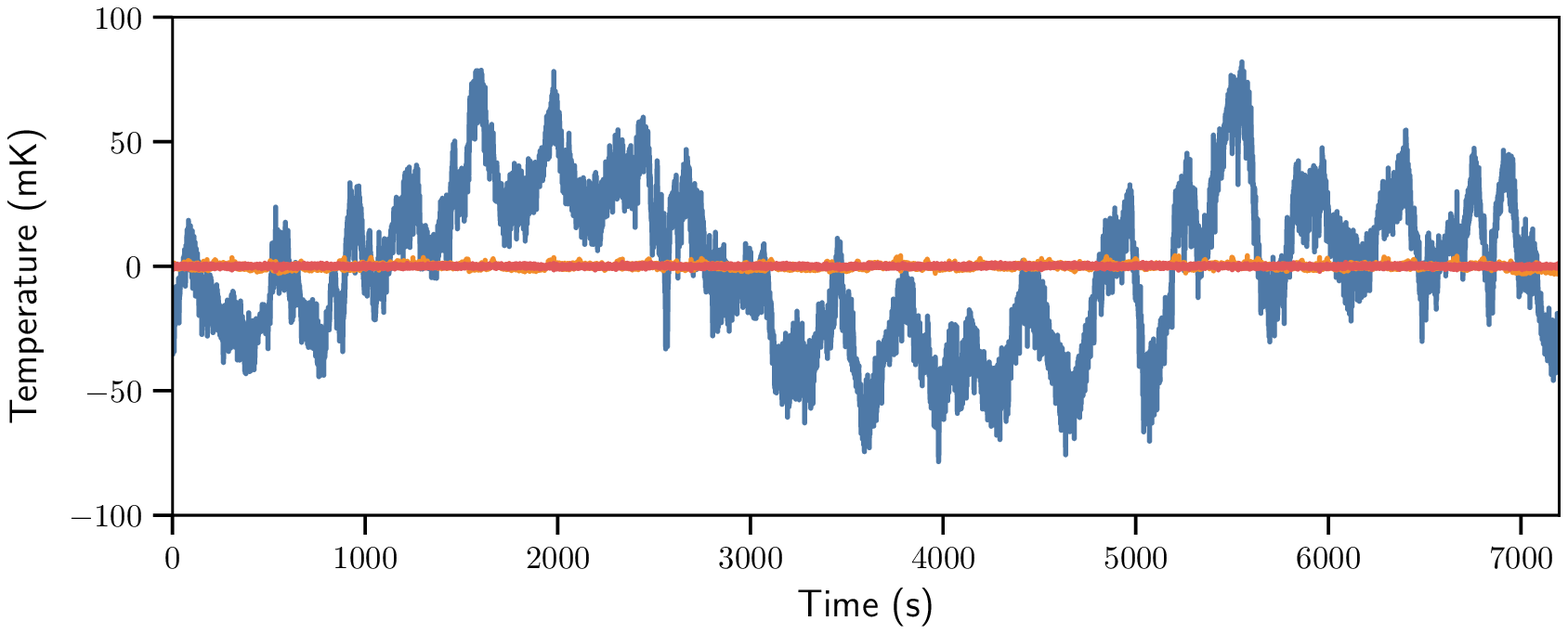}\\
         \includegraphics[width=0.95\textwidth]{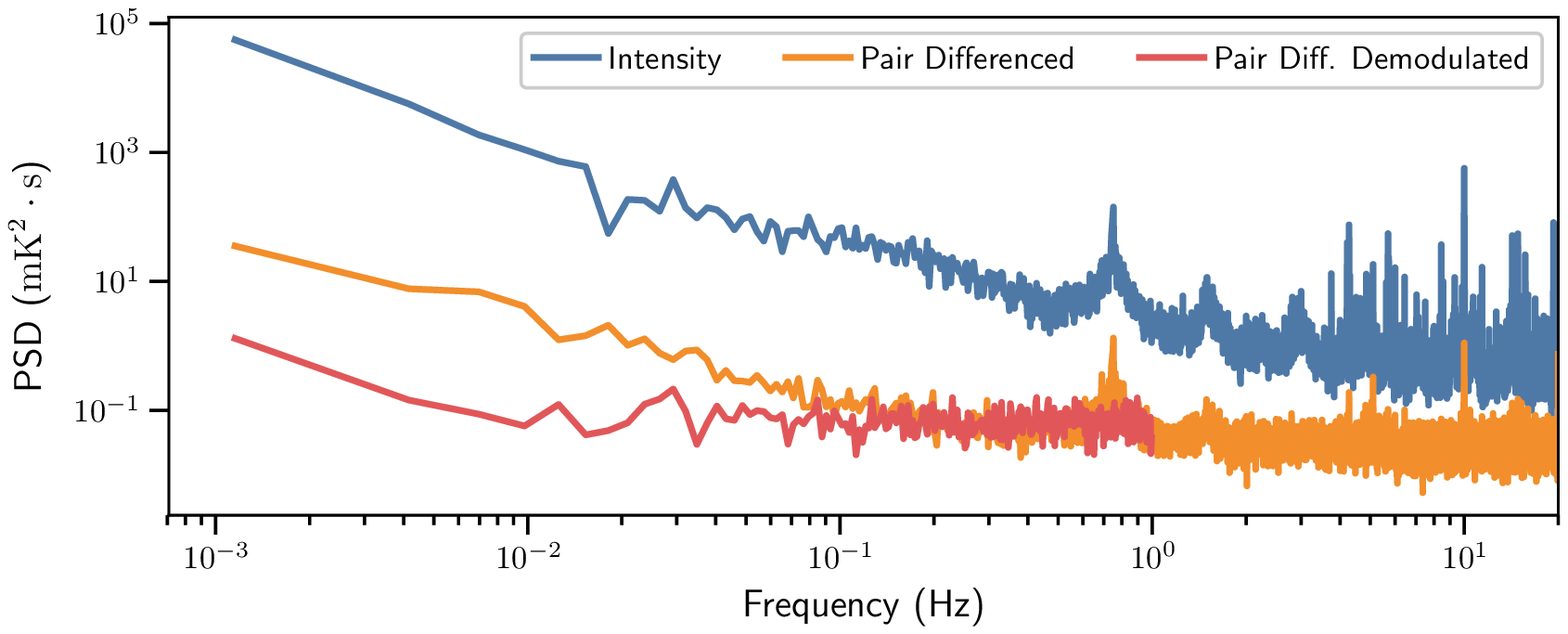}
    \end{tabular}
    \caption{\label{fig:ex_tod_psd} An example timestream (top) and power spectral density (bottom) for one of the 2-hour segments used in this analysis. The intensity, plotted in blue, is the pair-added data from a single pixel. The same data are shown pair-differenced in orange and pair-differenced then demodulated in red. The intensity timestream shows long timescale drifts and an azimuth synchronous signal. The intensity timestreams show several line features from the instrument, which are described in the text. The pair-differenced data are a measure of how CLASS would do without a modulator, and thus the level at which scan modulation is possible for non-modulated small aperture experiments from the Atacama. The pair-differenced data have knee frequencies between 0.1--1~Hz. Demodulation increases the stability further, with knee frequencies that are about a factor of 20 times smaller than pair-differencing alone. The demodulated white noise level is higher than the pair-differenced white noise by a factor of about $1/\epsilon_\mathrm{mod}$ (see Section~\ref{sec:mod_eff}), which is the effect of the linear polarization modulation efficiency for the VPM.}
\end{figure*}

The processing for these segments is similar to that implemented for other CLASS analyses. Glitches and jumps in the data due to transient detector effects are repaired by fitting a quadratic function to each side of the discontinuity, aligning the two functions, and interpolating in the cut region. For this study, a model for the VSS and a noise estimate are added to the cut region to smoothly join the two regions and enable FFT-based periodogram estimation of the power spectral density of the timestream. Next, the Butterworth filter applied to the data by the warm readout electronics is deconvolved.


As first described in \cite{appe19}, the unique oscillatory nature of the VPM enables one to utilize the VSS to measure the time constants associated with the detector response. For computational speed and robustness, each segment is split into 10-minute sections. A single-pole filter model that optimally removes hysteresis from the VSS at the mirror turn-around is fit to each section of data. The median of this data set is then deconvolved from the entire 2-hour segment. After time constant deconvolution, the timestreams are converted to Rayleigh–Jeans temperature units using the methods and values described in \cite{appe19}. 

The data at this point in the processing will be referred to as ``single detector data'' and, assuming $I_\mathrm{in}$ is nearly entirely unpolarized, this timestream takes the form 

\begin{equation}
\begin{aligned}
    d_{\pm} = (1\pm\lambda_{P,\mathrm{inst}}) I_\mathrm{in} &+ (P_\mathrm{in}+\lambda_P I_\mathrm{in}) S_P(z)  \\
     &+ (V_\mathrm{in}+\lambda_V I_\mathrm{in}) S_V(z)\\
     & + T_\mathrm{VPM} \, A_{z}(z).
\end{aligned}
\label{eq:single_det_data}
\end{equation}

\noindent The $\pm$ denotes the difference between orthogonal detector pairs and $A_z$ is the VSS, which has amplitudes {$\sim 55-110$~mK} in single detector data. The $\lambda$'s denote the leakage of intensity into different aspects of the signal. $\lambda_{P,\mathrm{inst}}$ is the temperature to polarization ($T\rightarrow P$) leakage that occurs after the sky signal is modulated by the VPM. $\lambda_{P}$ and $\lambda_{V}$ are the $T\rightarrow P$ leakages modulated by the VPM into linear and circular polarization, respectively. The first term in Equation \ref{eq:single_det_data} is the only term that does not depend on the VPM position and is therefore the only factor outside the 10~Hz VPM modulation band. Pairs of detectors are co-added to get the total intensity signal, shown in blue in Figure \ref{fig:ex_tod_psd}. A 1~Hz lowpass filter has been applied to all the timestreams in the top plot of Figure \ref{fig:ex_tod_psd} to remove the modulated components.

The single detector data are pair-differenced to create a timestream of the form
\begin{equation}
\begin{aligned}
    d_{pd} = \lambda_{P,\mathrm{inst}} I_\mathrm{in} & + (P_\mathrm{in}+\lambda_P I_\mathrm{in}) S_{P,pd}(z)  \\
     & + (V_\mathrm{in}+\lambda_V I_\mathrm{in}) S_{V,pd}(z)\\
     & + T_\mathrm{VPM} \, A_{\pm}(z),
\end{aligned}
\label{eq:real_pd}
\end{equation}
\noindent where the $S_{i,pd}$'s are the modulation functions accounting for pair-differencing and $A_{\pm}(z)$ is the difference in the VSS signal between the two orthogonal detectors. $A_{\pm}(z)$ has amplitudes ranging from {$\sim 12-55$~mK} depending on position in the focal plane. An example of a pair-differenced timestream is shown in orange in the top plot of Figure \ref{fig:ex_tod_psd}. Since polarization incident on the VPM is nearly entirely modulated, outside the 10~Hz modulation band the pair-differenced data are dominated by the instrument leakage of temperature into polarization that occurs after the VPM. The analysis of CLASS 40~GHz data is based on pair-differenced data because of the correlated noise between paired orthogonal detectors reported in \cite{appe19}.

The pair-differenced data are demodulated following the process described in Section \ref{sec:demod}. The resulting timestreams have the forms
\begin{equation}
    P = P_\mathrm{in}+\lambda_P I_\mathrm{in} + \langle S_{P,pd} \cdot A_{\pm}\rangle T_\mathrm{VPM}
    \label{eq:real_p_demod}
\end{equation}
\noindent and
\begin{equation}
    V = V_\mathrm{in}+\lambda_V I_\mathrm{in} + \langle S_{V,pd} \cdot A_{\pm}\rangle T_\mathrm{VPM},
    \label{eq:real_v_demod}
\end{equation}
\noindent where $\langle S_{P,pd} \cdot A_{\pm}\rangle$ and $\langle S_{V,pd} \cdot A_{\pm}\rangle$ represent the projection of the residual VSS into the linear and circular polarizations, respectively.

\subsection{Temperature to Polarization Leakage}

A measurement of the $T\rightarrow P$ leakage present in the system is done by comparing the results of the different data timestreams described above. All timestreams are low-passed with a 1~Hz filter and down-sampled to isolate the longer time scale modes in the system. Polarized azimuth synchronous ground pickup or sky signal could bias estimates of the $T\rightarrow P$ leakage in the timestreams, since distinguishing between the two requires much more than 2~hours of data. An azimuth synchronous template was fit to each timestream in each of the four data sets (single detector, pair-differenced, demodulated linear, and demodulated circular) and removed from the data. The template was created using fits to the first 19 terms of a Fourier series, a method chosen to limit the scope of what was removed to known frequencies since the azimuth synchronous signal is often below the noise in the demodulated data. This method will also remove the atmospheric circular polarization, due to Zeeman splitting of oxygen molecules in Earth's magnetic field, which was measured by \cite{petr20}. 

The different $T\rightarrow P$ leakage terms in Equations \ref{eq:real_pd}, \ref{eq:real_p_demod}, and \ref{eq:real_v_demod} are estimated using template fits to the pair-added intensity, $I$, for each set of data. The leakage terms ($\lambda_{P,\mathrm{inst}}$, $\lambda_{P}$, and $\lambda_{V}$) for each set of timestreams were calculated as

\begin{equation}
    \lambda_x = \frac{ I\cdot x}{ I\cdot I },
\end{equation}

\noindent where $x$ is the pair-differenced ($d_\mathrm{pd}$), demodulated linear ($P$), or demodulated circular ($V$) timestreams. Each individual measurement of $\lambda_x$ is low signal-to-noise given the small level of $T\rightarrow P$ leakage expected with a front-end polarization modulator: the formal statistical errors are $\sim0.004$ per measurement, while the expected signals are $\lesssim 10^{-4} - 10^{-5}$. These measurements are performed on all \nTotalTimestreams timestreams to build a probability distribution describing these effects.

\subsection{\label{sec:method_psd}Power Spectrum Fits}

Power spectral densities (PSDs) are calculated for the timestreams in each of the four data sets described in the previous section. As with the $T\rightarrow P$ leakage measurements, a Fourier series fit was used to create an azimuth synchronous signal template, which was subtracted from the timestreams before the Fourier transforms were performed. This removed lines in the PSDs at harmonics of the scan frequency. Examples of these PSDs are shown in the bottom frame of Figure \ref{fig:ex_tod_psd}. The PSDs for intensity show several features, such as the spacing of the teeth in the azimuth gear at 0.75~Hz and the pulse tube cryo-cooler frequency at 1.4~Hz. The collection of lines at 5~Hz, 10~Hz, and 15~Hz are due to the VSS signal; the 5~Hz and 15~Hz portions of the VSS signal are due to a design flaw in the 40~GHz VPM, which was corrected in the other CLASS VPMs \citep{harr18}. Figure \ref{fig:ex_tod_psd} shows the dramatic drop in low frequency noise achieved through demodulation.

With the exception of the few line features in the non-demodulated data, these PSDs are well defined by a white noise level, $\omega_n$, and a single $1/f$ term, parameterized with a knee frequency and spectral index, that describes the long timescale noise in the system:

\begin{equation}
    P(f) = w_n^2 \left(1+\left( \frac{f}{f_\mathrm{knee}} \right)^\alpha \right).
    \label{eq:one_over_f}
\end{equation}

\noindent The PSDs of all timestreams were fit to Equation \ref{eq:one_over_f} with frequency masks used to remove the noise lines present in the non-demodulated data. The $1/f$ knee frequencies of the demodulated data are very low ($\lesssim 20$~mHz). Even when using 2~hours of data, the majority of the $1/f$ section of the spectrum occurs in the lowest and least sampled frequency bins. Therefore, fits for the knee frequency and spectral index must account for the lower sampling and higher variance in these bins. The data were binned in frequency space, and the model was fit in log-space via the algorithm presented by \cite{Papadakis1993}; this method accounts for the sampling difference in the lower frequencies.

\begin{figure}
    \centering
    \includegraphics[width=\columnwidth]{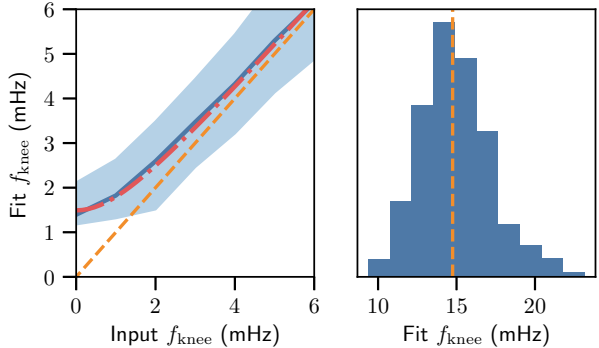}
    \caption{The results of running the PSD analysis on synthetic timestreams with known $1/f$ behavior. The left plot shows the median fit knee frequencies as a function of input knee frequency. The shaded area denotes the region that contains 68\% of the fits. The median fit knee frequencies are biased at the lowest frequencies due to the windowing, assumption of $1/f$ behavior, and the low levels of statistical sampling of the lowest frequency bins in a PSD. The red dot-dashed line shows a hyperbolic function fit to fit knee frequencies. This function is used to de-bias the knee frequencies in the datasets that contain a significant number of low knee frequencies. The right plot contains a histogram of the fit knee frequencies for timestreams with a known 15~mHz knee frequency. This indicates that for fits around 15~mHz, individual measurements have uncertainties (68\% confidence) of $\sim 2$~mHz.}
    \label{fig:fit_bias}
\end{figure}

Simulated timestreams were used to test for bias in the fitting procedure and to estimate the uncertainty in the resulting fit parameters.  First, timestreams were simulated with a spectral index of $-1.5$ and range of values for $1/f$ knee frequencies, with 1000 timestreams at each input value. The synthetic data were then run through the same analysis and fitting procedure as for the real data. The fit knee frequencies are shown in Figure \ref{fig:fit_bias}. The solid lines and shaded bands show the median and 68\% range of the distribution at each frequency respectively. The median and percentile bands are used because the fit distributions are non-Gaussian. Inputting white noise (``$f_\mathrm{knee} = $ 0 mHz'') results in an output knee frequency of $1.39^{+0.72}_{-0.22}$~mHz, indicating the level at which the windowing and assumption of a knee frequency bias the fit. For a 3~mHz input knee frequency the fit knee is $3.47_{-1.01}^{+0.97}$~mHz and for a 5~mHz input knee frequency the fit knee is $5.30_{-1.16}^{+1.31}$~mHz, indicating the bias is significantly less around these values. For the demodulated linear data set, fit knee frequencies below these values represent less than 1.3 and 4.0\% of the data, respectively. For the values most common in the demodulated linear data set, $7 \lesssim f_{\mathrm{knee}} \lesssim 20$~mHz, the average bias is $\sim1\%$. 

The bias is more significant in the demodulated circular and stare data sets, where 47\% and 31\% of the fit knee frequencies are below 5~mHz, respectively. For these datasets, a hyperbolic function of the form $f_\mathrm{fit} = \sqrt{f_\mathrm{true}^2+A}$ is fit to the bias curve. The inverse of this function is used to de-bias the circular and stare datasets; fits with knees below the $y$-intercept ($\sqrt{A}$) are assumed to be indistinguishable from white noise on these timescales. This debiasing results in an almost 10\% decrease in the median knee frequencies for circular polarization. In comparison, this technique would produce a 1\% decrease in the medians of the linear polarization knee frequencies.

The fits to synthetic timestreams can also be used to estimate the uncertainty for each individual fit to the data. The 68\% widths of the distributions for the range covering the majority of the demodulated data set are $10-20$\% of the median value. An example of this is shown in the right plot of Figure \ref{fig:fit_bias}. An individual fit with a knee frequency of 15~mHz can be expected to have $\sim2$~mHz errorbars. The uncertainty on the spectral index increases with decreasing knee frequency; it generally has {$10-20$\%} errors over the range of interest for the demodulated linear data.  The widths of these distributions require that many measurements are used to achieve a robust result for a measurement of the long timescale noise in the system. Conversely, the uncertainty on the white noise level is well-measured with only 1\% errorbars on the resulting values.


\section{Results}

\setlength{\extrarowheight}{4pt}
\begin{deluxetable*}{rcccc}[t!]
\tablecaption{\label{tab:demod_stats} The first three rows list median $1/f$ fit values for each distribution measured in this analysis, where errors indicate the 68\% widths of these distributions. The distribution widths are due to a combination of the precision of each measurement and observed variation within the data set. The bottom two rows list the 95\% confidence level of the temperature to polarization leakage across the focal plane. The calculation of these numbers is described in the text.}
\tablehead{\colhead{} & \colhead{} & \colhead{} & \multicolumn{2}{c}{Demodulated} \\ \colhead{} & \colhead{\hspace{7mm}Single Detector}\hspace{7mm} & \colhead{\hspace{7mm}Pair-Differenced}\hspace{7mm} & \colhead{\hspace{7mm}Linear}\hspace{7mm} & \colhead{\hspace{7mm}Circular}\hspace{7mm}}
\startdata
    Knee Frequency  & $3.29^{+2.98}_{-1.66}$~Hz & $350^{+611}_{-189}$~mHz & $15.12^{+21.6}_{-8.6}$~mHz & $4.9^{+10.7\dagger}_{-3.9}$~mHz \\
    Spectral Index  & $-1.53^{+0.29}_{-0.33}$ & $-1.62^{+0.30}_{-0.26}$ & $-1.46^{+0.34}_{-0.34}$ & $-1.44^{+0.45}_{-0.55}$\\
    White Noise ($\mathrm{\mu K_{RJ} \sqrt{s}}$)    & $393^{+172}_{-90}$ (NET) &  $231^{+25}_{-18}$~ (NEQ/U) & $339^{+52}_{-39}$ (NEQ/U) & $712^{+76}_{-60}$ (NEV) \\ 
    \hline
    $T \rightarrow P$ & & $|\lambda_\mathrm{P,inst}|< 8.0 \times 10^{-2}$ & $|\lambda_\mathrm{P}|<3.8\times 10^{-4}$ & \\
    $T \rightarrow V$ & & & & $|\lambda_\mathrm{V}|<9.1\times 10^{-4}$  \\
    \hline
\enddata
\tablecomments{All noise levels listed in this table are scaled to be ``per detector,'' meaning a factor of $\sqrt{N_\mathrm{det}}$ is used to calculate the noise level of the array for each measurement. Section \ref{sec:mod_eff} explains the expected differences between these values. \\ $^\dagger$The circular knee frequencies have been de-biased following the method in Section \ref{sec:method_psd}.}
\end{deluxetable*}

\begin{figure}
    \centering
    \includegraphics[width=\columnwidth]{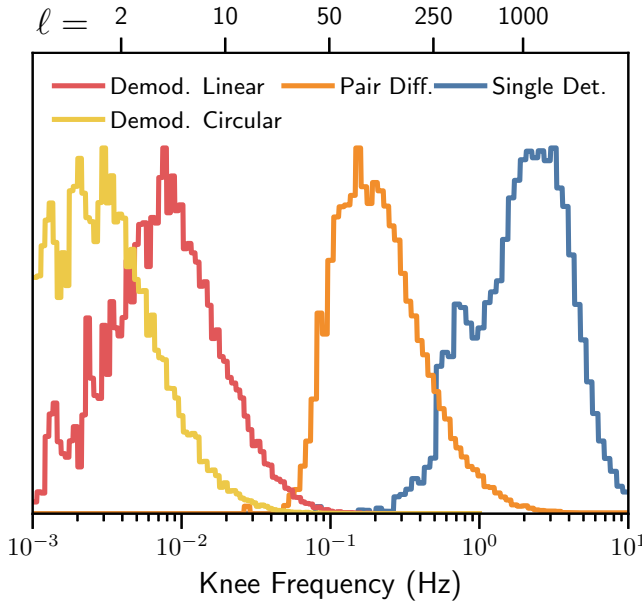}
    \caption{The knee frequencies fit for the single detector, pair-differenced, and pair-differenced demodulated data. The pair-differenced data, indicating what CLASS could be expected to achieve \textit{without} the front-end VPM, have knee frequencies about an order of magnitude lower than that of the single detector data. The pair-differenced demodulated knee frequencies, indicating the stability level CLASS is achieving, are an order of magnitude smaller than the pair-differenced data and two orders of magnitude lower than the single detector data. The shapes of these distributions are influenced by the uncertainty in individual fits and by the range of observing conditions. The multipole moments marked on the top of the plot match the observation frequency to the angular scale scanned on the sky for the CLASS scan strategy.}
    \label{fig:all_knees}
\end{figure}

The white noise, $1/f$ fits, and $T\rightarrow P$ measurements for each 2-hour long timestream are compiled into probability distributions that describe the variation within the data set and the uncertainty on the individual measurements. For the $1/f$ fit values, the medians and 68\% distribution widths for each variable and data set are compiled in Table \ref{tab:demod_stats}. These metrics are chosen because many of these distributions, especially for knee frequency, are significantly non-Gaussian. The knee frequencies and white noise distributions are in Figures \ref{fig:all_knees} and \ref{fig:noise_hist}, respectively. Since each individual $T\rightarrow P$ measurement is very low signal-to-noise, we look at the mean for each pixel, as $T\rightarrow P$ is not expected to change for a static optical setup. The absolute value of the per pixel $T\rightarrow P$ is what affects mapping and cosmological analyses; 95\% of the pixels have $T\rightarrow P$ leakages below the level listed in Table \ref{tab:demod_stats}. This excludes two poorly behaving pixels on the edge of the focal plane.\footnote{These pixels, compared to the rest of the focal plane, each have less than 1/3 the number of segments passing data quality cuts and will not be included in the larger CLASS 40~GHz analysis.}

While the main focus of this analysis is the demodulated data, the single detector data and pair-differenced data are also used to analyze the stability of the instrument and atmosphere for small aperture telescopes operating at 40~GHz in the Atacama desert. Figure \ref{fig:all_knees} shows the probability distributions for the knee frequencies of all four data sets in this analysis. The telescope scan strategy is used to convert the temporal based knee frequencies to the angular scales (multipole moments) covered by the telescope during that time. This conversion does not account for all aspects necessary to produce maps with sensitivity to a particular multipole moment of the CMB, but can be used as a reference for the stability level in the data. If $1/f$ knee frequencies are not sufficiently close to a targeted $\ell$ range, then the filtering required to map the dataset will also significantly reduce the sensitivity to that range. The exact requirements for knee frequency with respect to targeted $\ell$ range depend on the exact mapping technique.


The single detector knee frequencies in Figure \ref{fig:all_knees} illustrate why it is difficult to make CMB temperature maps of the larger angular scales ($\ell\lesssim50$) from the ground. Variations in atmospheric emission and instrument stability create $1/f$ noise that is much higher than the level required to observe the lowest multiple moments. The single detector data determine a starting point for instrument stability, but the CLASS analysis has focused on pair-differenced data because of the pixel-correlated noise in the focal plane \citep{appe19} and a better understanding of the pair-differenced polarized beams \citep{xu20}. Pair subtraction removes the first-order effects of many unpolarized sources of $1/f$. Unpolarized atmospheric emission, telescope temperature changes, and readout-column synchronous systematics are all significantly reduced by pair subtraction.  


The pair-differenced data are a measure of how CLASS would do without a modulator, and thus the level at which scan modulation is possible for non-modulated small aperture experiments from the Atacama. For CLASS 40~GHz pair-differenced data, 83\% of the timestreams have knee frequencies below 1~Hz, and the median knee frequency is 346~mHz. Since the VPM is placed at the front-end of the CLASS telescopes, the long timescale sections of the pair-differenced data contain a measurement of the $T\rightarrow P$ leakage of the instrument \textit{after} the VPM. These effects are unmodulated and will not affect the demodulated data. All sky and instrument polarization before the VPM is modulated up to the modulation frequency and is present in the demodulated pair-differenced data. The demodulated pair-differenced data are what CLASS actually achieves in terms of long term stability.

The pair-differenced demodulated linear polarization data, for the entire data set, have a median knee frequency of $15.12\pm0.15$~mHz, where the error is the estimated error on the median, found by bootstrapping. The bootstrap method calculates the standard deviation of the median values found when re-drawing the sample 1000~times. As will be discussed in Section \ref{sec:obs}, this knee frequency depends on a variety of instrumental and observational parameters such as PWV, boresight angle, and position in the focal plane. Naively mapping this knee to $\ell$ for the CLASS scan strategy gives $\ell_\mathrm{knee} \sim 7$, well into the targeted largest angular scales on the sky. We find the absolute value of the $T\rightarrow P$ leakage, $|\lambda_P|$, to be $<3.8\times 10^{-4}$ (95\% confidence), across the focal plane. This value is consistent with that measured in \cite{xu20} using beam mapping and is over two orders of magnitude lower than what is observed in the pair-differenced data before demodulation, emphasizing the critical advantage of the CLASS front-end polarization modulator.

The knee frequencies of the pair-differenced demodulated circular polarization data are much lower than for linear polarization. This is primarily due to a lower theoretical $T\rightarrow V$ level due to the VPM. The wire grid and mirror emission signature projects more closely to the linear modulation functions than the circular. The observed $T\rightarrow V$ limit is higher than the $T\rightarrow P$ limit due to a higher measurement uncertainty. As discussed in Section~\ref{sec:method_psd}, $\sim47\%$ of the circular knee frequencies are low enough to be biased by the $1/f$ fitting method and the results in Table~\ref{tab:demod_stats} are presented after de-biasing. 


An in-depth comparison between the CLASS 40~GHz $T\rightarrow P$ and knee frequency measurements to those of other CMB experiments with polarization modulators is difficult because of the subtleties involved in comparing instruments with different design goals. Every experiment that has implemented polarization modulation has used a different combination of observation frequency, scan pattern, beam width, modulator type, and modulator placement. ABS, which used a front-end continuously rotating half-wave plate (HWP) \citep{abs_hwp}, has the most similar optical design to the CLASS telescopes. ABS observed at 150~GHz and, motivated by a different targeted $\ell$ range, had a significantly higher scan frequency and observed smaller patches of sky. Beam maps and variations in the half-wave plate synchronous signal amplitude versus loading were used to constrain the ABS $T\rightarrow P$ leakage to $<7\times10^{-4}$ (95\% confidence) across the focal plane \citep{essinger2016}. While the CLASS 40~GHz $T\rightarrow P$ is measured at a similar level to ABS, we find knee frequencies significantly higher than their reported 2.0~mHz \citep{abs_hwp}, even after accounting for the difference in white noise levels. A similar result is found when comparing to continuous HWP tests by the \textsc{Polarbear} experiment \citep{Takakura2017}. As will be discussed in Section \ref{sec:obs}, we find this dichotomy to be explained by non-($T\rightarrow P$) related $1/f$ noise. 

The knee frequencies for pair-differenced data are an order of magnitude smaller than those for single detector data, and demodulating that data leads to another order of magnitude improvement in the long term stability of the data. The time required to integrate un-correlated noise scales with the square of the noise level. At 1~mHz, pair-differencing reduces this integration time by a factor of about 60; demodulating reduces the integration time by another factor of about 110. The stability and $T\rightarrow P$ leakage achieved with a front-end VPM is critical for CLASS to achieve the scientific objective of mapping the large angular scale CMB polarization from the ground. The results presented in Table~\ref{tab:demod_stats} contain no post-processing beyond the azimuth-synchronous fit removal described in Section~\ref{sec:methods}. Thus, these data represent a starting point in the mapping pipeline, where the removal of low-order polynomials or other destriping techniques can be used to further suppress the long time-scale noise.

\subsection{\label{sec:mod_eff} Modulation Efficiency}

\begin{figure}
    \centering
    \includegraphics[width=\columnwidth]{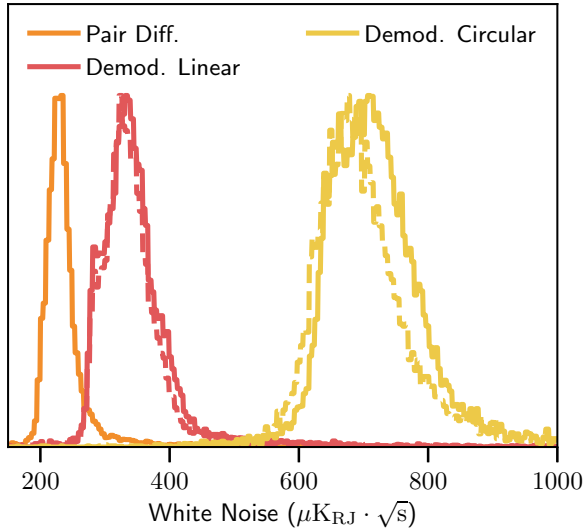}
    \caption{The white noise distributions of the pair-differenced and demodulated data highlighting the difference in modulation efficiency between the linear and circular demodulated data. The sinusoidal VPM throw is set to maximize sensitivity to the linear polarization. The dashed lines mark the noise levels predicted from the pair-differenced data and the theoretical modulation efficiency. The difference between the solid and dashed lines indicate a slightly higher than expected noise level for the circular polarization timestreams, possibly due to some unknown VPM synchronous noise source.}
    \label{fig:noise_hist}
\end{figure}

The modulation efficiency, $\epsilon_\mathrm{mod}$, of the system is the ratio of the demodulated to the pre-demodulation white noise level.\footnote{This was called $f_\mathrm{mod}$ in \cite{harr18}. It is changed here because we use $f_\mathrm{mod}$ to describe the modulation frequency.} For VPMs, as described in \cite{harr18} for linear polarization, the expected modulation efficiency depends primarily on the mirror throw parameters, the bandwidth of the detectors, and the beam incidence angle on the VPM. Given the detector bandpass and instrument optical design, the mirror throw parameters are set to maximize the linear modulation efficiency. 

\cite{harr18} Equation 4 defines a simpler method of calculating $\epsilon_\mathrm{mod}$, which only looks at the modulation function of the parameter in question (linear or circular polarization, $S_P$ or $S_V$). The more complete way to calculate the expected modulation efficiency is propagation of uncertainties through Equations \ref{eq:p_demod} and \ref{eq:v_demod}. This method, which accounts for the covariance between $P$ and $V$ in the modulation, differs from the simpler method at sub-percent levels. The more rigorous calculation is used for the results discussed here. Accounting for the actual positions of the VPM mirror during data taking and the spread in modulation efficiency due to the range of detector incidence on the VPM, the median modulation efficiency for this data set is expected to be $0.7001\pm0.0002$ for linear polarization and $0.3316\pm0.0001$ for circular polarization. The uncertainties are the estimated error on the median calculated using bootstrapping. The predicted values have 68\% distributions widths of about $\pm0.03$ due to the difference in modulation efficiency across the focal plane and accounting for the variation in VPM mirror throw during the timestreams.

Figure \ref{fig:noise_hist} shows histograms of the white noise levels for the pair-differenced, demodulated linear, and demodulated circular timestreams. All these values are normalized to be per detector instead of per feedhorn (or pixel), meaning the per feedhorn noise measurements are multiplied by a factor of $\sqrt{2}$ to obtain the per-detector noise figures reported here. The dashed lines denote the distributions predicted based on the pair-differenced data and the theoretical modulation efficiencies calculated from the VPM positions for each particular timestream. 

Both the linear and circular polarization timestreams have slightly higher white noise levels than would be expected from VPM demodulation alone. The observed median modulation efficiency for linear polarization is $0.689\pm0.001$ and for circular polarization it is $0.3247\pm0.0003$. This difference indicates the presence of a small level VPM synchronous noise. This is a 1\% effect for the linear polarization timestreams and a 3\% effect for the circular timestreams, small enough to have no significant impact on the sensitivity of the CLASS instrument.



\section{\label{sec:obs}Sources of Demodulated \lowercase{$1/f$} Noise}

The median demodulated knee frequency for the first observing era of the 40~GHz CLASS telescope was 15.12~mHz; however, as visible in Figure \ref{fig:all_knees}, the distribution of all knee frequencies for the demodulated data has a large width. The measured demodulated knee frequencies for each 2-hour segment were compared to various characteristics of the telescope and site during that time, such as the PWV, air temperature, wind speed, boresight angle, and mirror temperatures. Here we discuss the observational characteristics that have the largest effect on the knee frequencies of the demodulated data.

\setlength{\extrarowheight}{4pt}

\begin{deluxetable*}{lcccc}[t!]
\tablecaption{\label{tab:demod_stats_cuts} The median values of the $1/f$ spectra fit parameters for a variety of different demodulated linear-polarization data splits. The first uncertainty is the bootstrapped error on the median, and the bimodal uncertainties indicate the 68\% widths for each distribution. Unless otherwise specified, the data use only the central pixels as defined in the text. The ``epochs'' listed are time periods during Era~1 marked by changes in the instrument configuration. The Early Epoch was the initial configuration post commissioning, the Stiff Epoch is after stiffening measures were implemented on the telescope mount, and the Black Epoch is after absorptive blackening was applied to the forebaffle at the entrance of the telescope.}

\tablehead{\colhead{Data Split} & \colhead{$N_\mathrm{segments}$} & \colhead{Knee Frequency (mHz)} & \colhead{Spectral Index} & \colhead{White Noise ($\mathrm{\mu K_{RJ}} \sqrt{\mathrm{s}}$)} }
\startdata
All Pixels  &  15407  &  $15.12 \pm 0.15^{+21.25}_{-8.5}$  &  $-1.46 \pm 0.0^{+0.34}_{-0.34}$  &  $338.33 \pm 0.36^{+50.54}_{-38.58}$  \\
Edge Pixels  &  4264  &  $30.22 \pm 0.35^{+24.8}_{-14.57}$  &  $-1.49 \pm 0.01^{+0.28}_{-0.27}$  &  $357.17 \pm 0.83^{+56.21}_{-33.1}$  \\
Central Pixels  &  11143  &  $11.65 \pm 0.1^{+13.58}_{-6.03}$  &  $-1.45 \pm 0.0^{+0.35}_{-0.38}$  &  $331.11 \pm 0.42^{+43.55}_{-38.47}$  \\
\hline
PWV $<1.5$~mm  &  8712  &  $10.97 \pm 0.1^{+12.74}_{-5.68}$  &  $-1.44 \pm 0.01^{+0.37}_{-0.4}$  &  $329.62 \pm 0.48^{+44.33}_{-39.2}$  \\
PWV $>1.5$~mm  &  2431  &  $14.27 \pm 0.22^{+16.78}_{-6.93}$  &  $-1.45 \pm 0.01^{+0.3}_{-0.33}$  &  $336.97 \pm 0.94^{+39.86}_{-36.42}$  \\
\hline
Wind speed $<4$~m/s  &  8893  &  $11.26 \pm 0.11^{+13.11}_{-5.84}$  &  $-1.46 \pm 0.01^{+0.36}_{-0.4}$  &  $330.28 \pm 0.48^{+41.82}_{-38.02}$  \\
Wind speed $>4$~m/s  &  2250  &  $13.74 \pm 0.26^{+14.09}_{-7.21}$  &  $-1.4 \pm 0.01^{+0.33}_{-0.32}$  &  $334.44 \pm 1.04^{+52.18}_{-40.57}$  \\
\hline
BS $-45.0^\circ$  &  1772  &  $9.53 \pm 0.16^{+10.05}_{-4.58}$  &  $-1.55 \pm 0.01^{+0.34}_{-0.44}$  &  $331.05 \pm 1.01^{+31.15}_{-33.86}$  \\
BS $-30.0^\circ$  &  1695  &  $10.01 \pm 0.23^{+11.1}_{-5.25}$  &  $-1.56 \pm 0.01^{+0.37}_{-0.44}$  &  $335.32 \pm 1.02^{+45.17}_{-36.0}$  \\
BS $-15.0^\circ$  &  1439  &  $11.65 \pm 0.3^{+12.3}_{-6.32}$  &  $-1.46 \pm 0.01^{+0.34}_{-0.37}$  &  $332.68 \pm 1.31^{+38.29}_{-39.11}$  \\
BS $0.0^\circ$  &  1730  &  $12.64 \pm 0.26^{+14.95}_{-6.96}$  &  $-1.36 \pm 0.01^{+0.33}_{-0.36}$  &  $326.46 \pm 1.0^{+46.2}_{-38.07}$  \\
BS $+15.0^\circ$  &  1620  &  $13.32 \pm 0.29^{+14.25}_{-6.69}$  &  $-1.38 \pm 0.01^{+0.32}_{-0.32}$  &  $328.62 \pm 1.1^{+45.55}_{-40.52}$  \\
BS $+30.0^\circ$  &  1485  &  $13.18 \pm 0.32^{+15.85}_{-6.66}$  &  $-1.37 \pm 0.01^{+0.35}_{-0.38}$  &  $329.94 \pm 1.25^{+51.49}_{-40.3}$  \\
BS $+45.0^\circ$  &  1402  &  $12.39 \pm 0.28^{+16.67}_{-5.76}$  &  $-1.46 \pm 0.02^{+0.43}_{-0.35}$  &  $335.11 \pm 1.61^{+46.7}_{-42.85}$  \\
\hline
Early Epoch  &  2921  &  $13.82 \pm 0.24^{+15.31}_{-7.43}$  &  $-1.38 \pm 0.01^{+0.39}_{-0.37}$  &  $327.29 \pm 1.13^{+57.99}_{-38.85}$  \\
Stiff Epoch  &  1699  &  $10.98 \pm 0.26^{+11.5}_{-5.73}$  &  $-1.45 \pm 0.01^{+0.36}_{-0.32}$  &  $332.98 \pm 1.2^{+50.89}_{-38.12}$  \\
Black Epoch  &  6523  &  $10.99 \pm 0.12^{+13.06}_{-5.5}$  &  $-1.48 \pm 0.01^{+0.34}_{-0.41}$  &  $331.89 \pm 0.52^{+37.96}_{-37.14}$  \\
\hline
\multicolumn{5}{c}{Stare Data: From Second Observational Era} \\
\hline
All  &  139  &  $7.47 \pm 0.69^{+15.64\dagger}_{-4.25}$  &  $-1.74 \pm 0.03^{+0.31}_{-0.35}$  &  $335.57 \pm 2.31^{+47.12}_{-20.6}$  \\
Central Pixels  &  103  &  $6.95 \pm 0.8^{+15.29\dagger}_{-4.29}$  &  $-1.78 \pm 0.04^{+0.33}_{-0.31}$  &  $329.07 \pm 3.61^{+42.51}_{-19.18}$  \\
\hfill PWV $<1$~mm  &  50  &  $4.24 \pm 0.92^{+17.14\dagger}_{-2.86}$  &  $-1.77 \pm 0.07^{+0.41}_{-0.68}$  &  $322.39 \pm 3.39^{+42.47}_{-13.85}$  \\
\hfill PWV $>2$~mm  &  53  &  $8.46 \pm 0.82^{+12.53\dagger}_{-4.11}$  &  $-1.79 \pm 0.05^{+0.19}_{-0.19}$  &  $335.44 \pm 3.0^{+42.62}_{-16.24}$  \\
Edge Pixels  &  36  &  $8.92 \pm 2.56^{+17.09\dagger}_{-4.52}$  &  $-1.66 \pm 0.07^{+0.25}_{-0.45}$  &  $358.84 \pm 11.17^{+41.07}_{-24.31}$  \\
\hfill PWV $<1$~mm  &  19  &  $8.16 \pm 3.23^{+29.18\dagger}_{-3.75}$  &  $-1.52 \pm 0.05^{+0.16}_{-0.23}$  &  $375.03 \pm 18.6^{+67.03}_{-41.29}$  \\
\hfill PWV $>2$~mm  &  17  &  $10.03 \pm 3.92^{+16.28\dagger}_{-5.54}$  &  $-1.78 \pm 0.14^{+0.21}_{-0.43}$  &  $358.3 \pm 9.35^{+38.33}_{-18.06}$  \\
\enddata
\tablecomments{ $^\dagger$ These numbers have been ``debiased'' using the method described in Section \ref{sec:methods}.}
\end{deluxetable*}

\begin{figure}
    \centering
    \includegraphics[width=\columnwidth]{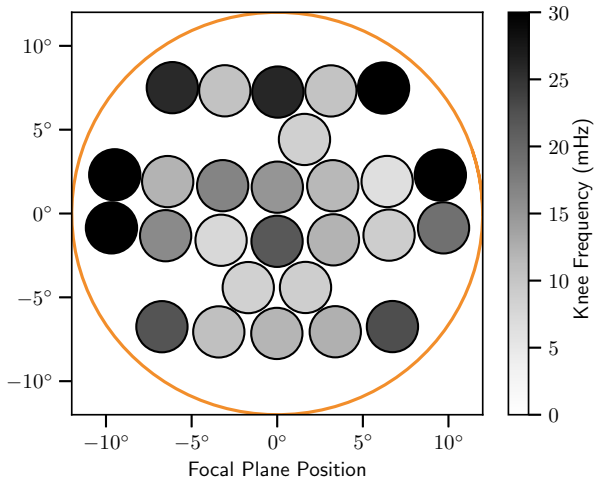}
    \caption{The median demodulated knee frequencies per pixel in the focal plane. The orange circle denotes the shape, but \textit{not} the actual radius, of the forebaffle at the entrance to the telescope. Eight pixels, the four left-most and the four right-most, have significantly higher demodulated knee frequencies than the more central pixels.}
    \label{fig:feed_knees}
\end{figure}

Table \ref{tab:demod_stats_cuts} lists the median knee frequency, spectral index, and white noise for a large variety of cuts on the demodulated linear polarization dataset. One of the data splits with the largest impact is the position in the focal plane. Figure \ref{fig:feed_knees} shows the median knee frequency for each pixel in the 40~GHz focal plane during the first observation era. The eight ``edge pixels,'' those closest to the orange circle in the figure, have beams that are closest to the forebaffle at the entrance to the telescope and have a median knee frequency about three times larger than all the other ``central pixels.'' \cite{xu20} found these pixels have features in the polarized beams consistent with a higher $T\rightarrow P$ leakage and, while included in the $<3.8\times10^{-4}$ constraint, this analysis also found higher $T\rightarrow P$ leakage in those pixels. The forebaffle is the part of the telescope before the VPM, meaning instrumental polarization ($T\rightarrow P$ leakage) or a spill, that changes as a function of VPM position could influence the demodulated timestreams. The size and shape of the forebaffle also influences the level of ground pickup. Any of these effects could cause increased $T\rightarrow P$ leakage or $1/f$ in the demodulated data. 

Unlike the polarized beams in \cite{xu20}, the higher knee frequency effect in the edge pixels was not changed by the installation of absorptive blackening on the forebaffle surface in July 2017 (denoted as the ``Black'' epoch in Table \ref{tab:demod_stats_cuts}). All other analysis in this section is presented without measurements from the edge pixels in order to isolate the investigation of the different sources of $1/f$. The forebaffle encroachment radius was increased and its shape was made more ovular during later observing seasons to accommodate two telescopes on the same mount. This change should have reduced the overall effect of the forebaffle and will affect how pixels on opposite sides of the focal plane interact with it. The stare data set, comprised of data from this second observational era, shows no statistically significant difference between the edge and central pixels. This is an indication that the forebaffle changes improved the performance of the telescope, but more analysis is necessary to verify this conclusion. 


When the knee frequencies are binned with respect to wind speed there is a distinct shift in median knee frequency at about 4.25~m/s. While we could expect the atmosphere to change faster with higher wind speeds, and thus increase the atmospheric contribution to the $1/f$, this would be a more gradually evolving effect. A more likely explanation of this shift is an excitation of a vibrational mode in the telescope structure. The rest of the analysis in this section is presented with observations for which the wind speed is less than 4 m/s.

\subsection{Red Noise Amplitudes}
The two observational characteristics with a significant and interconnected impact on the long term stability of the demodulated timestreams are PWV and boresight angle. Quantifying the contributions from different observational factors requires assuming multiple sources of $1/f$ noise exist in the demodulated timestreams. However, as described in Section \ref{sec:methods}, the spectra of individual PSDs are well described by only one $1/f$ term. This implies that either the $1/f$ noise sources have similar spectral indexes or some contributions are significantly smaller than others. Either case would make fitting to a more complex noise model difficult, so we will look at how the fit to a single $1/f$ noise source would be affected by the presence of multiple smaller noise sources. 

We will assume each $1/f$ noise source has the same $\alpha=-1.46$ for the entire data set. While not completely correct, this is a reasonable assumption since the large majority of $1/f$ sources have pink spectra with $-2 \leq \alpha \leq 0$ and we do not see large deviations from this when looking at different subsections of the dataset. We also note that while knee frequencies are generally defined with respect to a white noise level, the amplitude of the $1/f$ portion of the PSD can be referenced to any chosen pivot frequency. For this part of the analysis we choose

\begin{equation}
    f_\mathrm{p}^{-\alpha} = (10\text{ mHz})^{1.46}
\end{equation}

\noindent to convert knee frequency, $f_\mathrm{k,\mathrm{fit}}$, and white noise, $\omega_n$, to a ``red noise amplitude,'' $\omega_{r,\mathrm{fit}}$, at the pivot frequency. This noise amplitude is assumed to include various noise sources $\omega_s(t_\mathrm{obs})$ that vary with time and a baseline noise term $\omega_{r,\mathrm{base}}$ that accounts for the other $1/f$ noise not explicitly included in the source list:

\begin{equation}
\begin{aligned}
    \omega_n^2 f_{k,\mathrm{fit}}^{-\alpha} 
    &= \omega_{r,\mathrm{fit}}^2 f_\mathrm{p}^{-\alpha} \\
    &= \omega_{r,\mathrm{base}}^2 f_\mathrm{p}^{-\alpha}
    + \sum_{ s\in \mathrm{Sources}} \omega_\mathrm{r,s}^2(t_\mathrm{obs}) f_\mathrm{p}^{-\alpha}.
\end{aligned}
\end{equation}

\noindent Specifying the noise sources against the same pivot frequency enables direct comparisons of their amplitudes. The different sources of $1/f$ noise add in quadrature. 

\subsection{PWV Dependence}
\begin{figure*}
    \centering
    \includegraphics[width=0.95\textwidth]{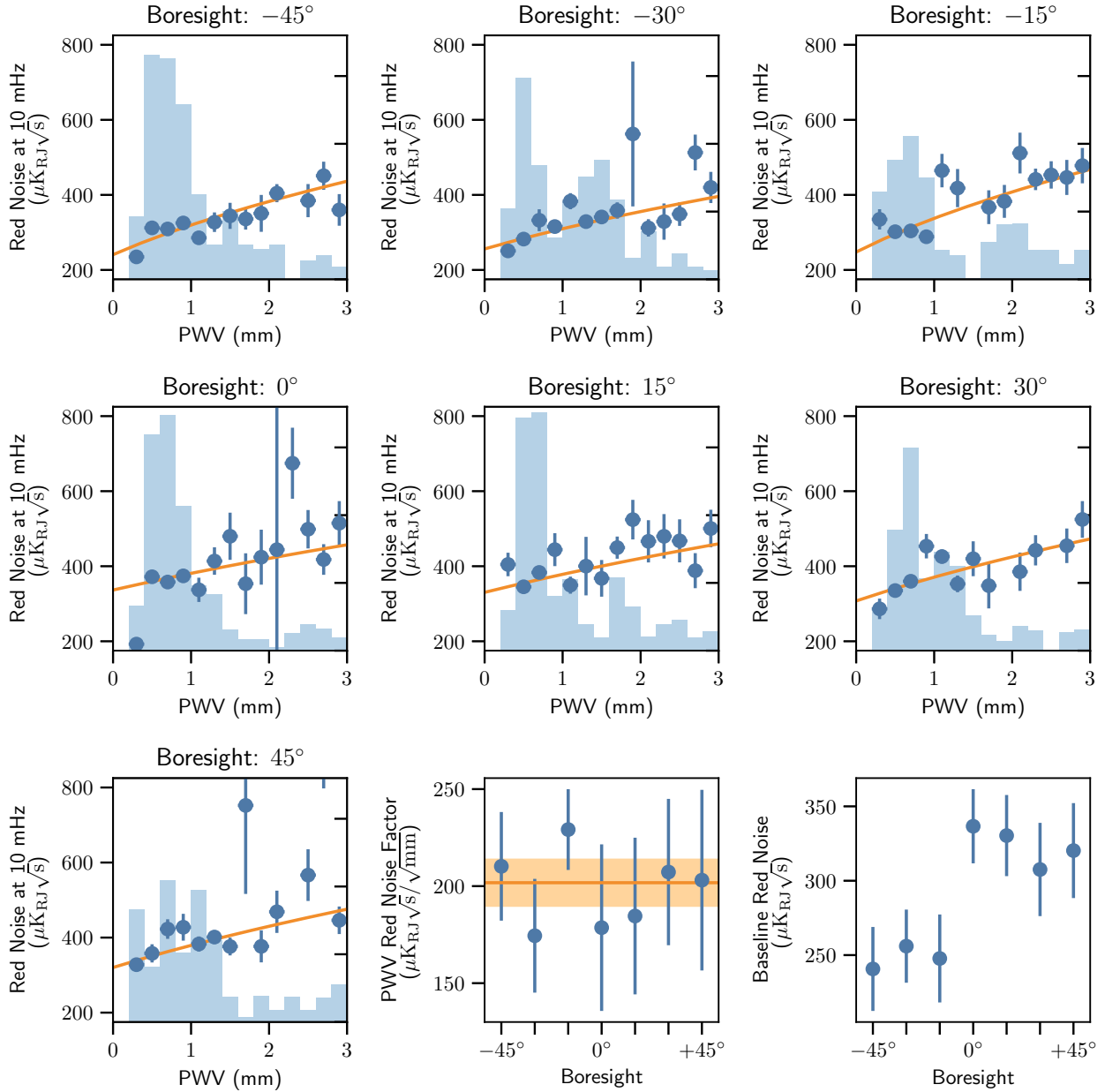}
    \caption{The first seven plots show the median red noise amplitudes at 10~mHz for each boresight angle when binned in PWV with uncertainties that are bootstrapped-estimated errors on the median. The median is used to prevent long tails in the distributions from significantly influencing the results. The lighter blue bars indicate the relative number of measurements in each bin. The orange lines are fits to the square of the red noise amplitude vs PWV for each boresight. The bottom middle (right) plot shows the slope ($y$-intercept) of the fits. The errors on the fit parameters come from bootstrapping the data used for each fit.}
    \label{fig:pwv_bore_trends}
\end{figure*}

 Figure \ref{fig:pwv_bore_trends} shows the median red noise amplitude of the data when split by boresight angle and PWV\footnote{We note that these PWV values, as measured by the APEX or ALMA experiments, are the values for PWV at Zenith, while the CLASS observations are at $45^\circ$ elevation.} in the first seven plots. The blue data points are the median red noise amplitude in that bin, and the errors are estimated using bootstrapping. The histograms in the background show the number of samples in each bin. The square of the red noise amplitude with respect to PWV is fit to lines for each boresight angle. The lowest PWV point for $0^\circ$ boresight is excluded from the fit because its value and uncertainty are abnormally low, which appears to be due to timing of when data in that point were taken. These fits are plotted in Figure \ref{fig:pwv_bore_trends} and have reduced $\chi^2$ values between 1.2 and 2.9, indicating the model fairly represents those data. The slopes and $y$-intercepts of these fits are plotted in bottom-middle and bottom-right plot, respectively, with errors estimated using bootstrapping. The fit values for the slope with PWV are not significantly different for the different boresight angles and their weighted average gives a measurement of the amplitude of the PWV related red noise in the demodulated timestreams,
\begin{equation}
    \omega^2_{r,\mathrm{pwv}} = (203 \pm 12)^2 \left( \frac{\mathrm{PWV}}{1~\mathrm{mm}} \right) \mathrm{\mu K_{RJ}^2 \cdot s}.
\label{eq:pwv_trend}
\end{equation}

\noindent The PWV-scaling can be subtracted, in quadrature, from the red noise amplitudes of every fit to the demodulated timestream PSDs. This results in a $\sim32\%$ reduction in median red noise across the data set. 


PWV tracks the amount of water vapor suspended in the atmosphere. Increasing PWV increases the overall thermal emission from the atmosphere and a relation between the amplitude of turbulence-sourced $1/f$ noise and PWV has been observed by large-aperture experiments \citep{LayHalverson2000, Dunner2013}. As discussed in Appendix \ref{sec:single_dets}, the CLASS 40~GHz single-detector timestreams do not observe the same $1/f$ spectrum as the large-aperture experiments, but it is not surprising that an atmospheric signal is observed in the demodulated data.

Following Equation \ref{eq:real_p_demod}, there are three mechanisms for how large-scale atmospheric fluctuations could enter into the demodulated timestreams. First, we could be observing an actual large-scale polarized signal from the atmosphere ($P_\mathrm{in} \neq 0$). CLASS has observed this in $V$, where the signal is due to Zeeman splitting of oxygen lines in the Earth's magnetic field \citep{petr20}; however the Zeeman effect produces negligible linear polarization and other sources of polarized molecular emission are not coherent on atmospheric length scales \citep{Hanany2003}.

\cite{Takakura2019} recently found evidence for polarized ice clouds in the upper atmosphere, where the polarization is almost entirely aligned along Stokes $Q$ in horizontal coordinates. The signal from these clouds can be quite significant ($\geq 200$~mK) at 150~GHz. However, scaling to 40~GHz and accounting for the CLASS beam size implies the signal from these clouds would be less than 1~mK for even the largest clouds. This level of signal is small enough to not be flagged in the data timestreams and could influence the $1/f$ spectrum when observing Stokes $Q$. If the polarized ice clouds were influencing the $1/f$ spectrum of the demodulated pair-differenced timestreams, the effect would be symmetric with boresight and the highest red noise amplitudes would be at $\pm45^\circ$ boresight angle, where instrument $U$ is aligned with horizontal $Q$. The pattern observed with boresight, shown in the bottom right plot of Figure~\ref{fig:pwv_bore_trends}, does not reflect what would be expected if polarized clouds were influencing the $1/f$ portion of the timestream PSDs.

Another unlikely mechanism is that we are observing atmospheric temperature fluctuations occurring at the VPM frequency. Effectively, anything with functional form $A$ where $\langle S_P \cdot A\rangle \neq 0$ could leak into the demodulation. This is very unlikely to happen for atmospheric fluctuations. They would be suppressed by pair-differencing and, given the $1.6^\circ$~FWHM of the beam and the $1^\circ/\text{s}$ scan frequency, atmospheric fluctuations should not be visible above $\sim400$~mHz; much lower than the 10~Hz modulation frequency. 

The last and most likely scenario is that large-scale atmospheric temperature variations could be leaking into the data through $T \rightarrow P$ leakage ($\lambda_P \neq 0$). The level of $T\rightarrow P$ leakage required to have the effect of Equation~\ref{eq:pwv_trend} on the demodulated data is difficult to estimate because we do not see similar patterns with boresight, PWV, and red noise amplitude for our single detector data. In addition, large-scale atmospheric contributions to the demodulated PSDs could be in the form of a stochastic $1/f$ spectrum or a set of slowly varying line features at harmonics of the scan frequency. The residual line features, after subtraction of a single azimuth template for the timestream, could bias the $1/f$ fitting to a higher knee frequency. The amplitudes of these effects are small; for this analysis, it is not possible to distinguish between these large-scale slow varying atmospheric contributions and slowly varying azimuth synchronous signal. This will require multi-frequency simultaneous observations from multiple telescopes and should be possible to test in the future, with CLASS, using simultaneous observations from four frequency bands. 

\subsection{Boresight Dependence}

While unlikely to be sourced by polarized clouds, the discrepancy between the positive and negative boresights is statistically significant. As shown in the bottom-right plot of Figure \ref{fig:pwv_bore_trends}, the baseline red noise ($\omega_{r,\mathrm{base}}$) goes from $248\pm15~\mathrm{\mu K_{RJ} \cdot \sqrt{s}}$ for negative boresights to $321\pm14~\mathrm{\mu K_{RJ} \cdot \sqrt{s}}$ for positive boresights. Neither range has a statistically significant trend as a function of boresight. This effect is likely to be caused by some type of mechanical vibration induced in the telescope structure. In the CLASS mechanical design, $0^\circ$ boresight is where many beams and joints transition  from being loaded under compression to under tension or vice versa, meaning a change in the resonance frequencies of the structure could be expected at that point. The influence of vibrations of either the focal plane or the optical elements could leak into the demodulated data. If a vibration induced signal, $A_V$, has a structure such that $\langle S_P \cdot A_V\rangle$ varies slowly with time, this will produce demodulated $1/f$. 

There are other factors which point to slowly varying vibration amplitudes as the source of the boresight dependence in the demodulated $1/f$ knees. First, the measured $T\rightarrow P$ leakage does not see an analogous pattern to the knee frequencies when split by boresight. In addition, there was a change when additional vibration suppression measures were implemented in December~2016 (denoted as the ``Stiff" epoch in Table \ref{tab:demod_stats_cuts}). The median baseline red noise for the negative boresights decreased by $105\pm9~\mathrm{\mu K_{RJ} \cdot \sqrt{s}}$ after these measures were implemented while the positive boresights decreased by $50\pm13~\mathrm{\mu K_{RJ} \cdot \sqrt{s}}$. Conversely, the baseline red noise levels per boresight differ by less than $1\sigma$ before and after absorptive blackening was applied to the forebaffle.


It is clear there are residual boresight dependent, VPM synchronous vibrations that are affecting the long-term stability of the demodulated data during the first observing era for the CLASS 40~GHz telescope. During this observing time, there was only one telescope on the telescope mount that was designed to hold two. Counterweights were installed on the opposite side, to partially replicate the weight of the uninstalled telescope. It is possible an uneven mass distribution contributed to these boresight dependent vibrations. Since the first observing era, the second telescope has been installed on that mount, and a third telescope was installed on a second mount. Preliminary analysis of Era~2 data does not show the same discontinuity with boresight, suggesting the evening out of the weight distribution or one of the other upgrades remidied this discontinuity.

The offset at positive boresights is subtracted, in quadrature, from the red noise amplitudes of every fit. The reduced red noise amplitudes are then scaled back to a knee frequency using the white noise level of the timestream. Adjusting for the positive boresight vibrational offset reduces the knee frequency of the central pixels from $11.65\pm0.1$ to $10.4\pm0.2$~mHz, corresponding to a $15\%$ reduction in the red noise amplitude. 

\subsection{Comparison with Stare Data}

A similar $1/f$ noise analysis was performed on the six stare observations during the second observing era for the 40~GHz telescope. These data, taken while the mount was not scanning, were used to get an estimate of the baseline level of $1/f$ possible with a front-end VPM. The data are processed identically as the scanning data and the fits to the PSDs are also listed at the bottom of Table \ref{tab:demod_stats_cuts}. The spectral indexes of these data are steeper than those of the scanning data, possibly indicating the $1/f$ noise for stare data is dominated by a source different than what dominates the scanning data.

The median knee frequency for stare data from the central detectors with PWV~$<1$~mm is $4.24\pm0.92$~mHz. This level of $1/f$ noise includes any variation in demodulated data due to the stability of the VPM alone and would also detect any fluctuations due to instabilities in the readout system if they were a dominating factor of the residual $1/f$. The estimated knee frequency of the scanning data-set when both the effects of PWV and boresight are removed is $7.5\pm2.0$~mHz; about 2$\sigma$ higher than the stare $1/f$ knee frequency.



Since intrinsic variability of the VPM does not account for the residual $1/f$ noise, once the boresight offset and PWV variation are removed, we are left to conclude the residual $1/f$ noise is due to something associated with scanning. 


\section{Conclusions} 

In this paper we have examined the long time-scale stability of the data from the 40~GHz CLASS telescope using a front-end variable-delay polarization modulator and found a substantial decrease in temperature to polarization leakage and knee frequencies between the pair-differenced and demodulated timestreams. This will enable CLASS to observe the largest-angular scales of the CMB polarization from the ground. 

We have presented a demodulation scheme that accounts for the covariance between the modulated linear and circular polarizations through a continuous least-squares fitting approach and outputs timestreams of the linear and circular polarization incident on the VPM as well as their associated covariances.

The demodulation process was applied to a selection of data taken by the 40~GHz CLASS telescope during its first observational era from September~2016 to March~2018. The ``raw'' single detector, pair-differenced, and pair-differenced demodulated data are fit to $1/f$ spectra and the $T\rightarrow P$ leakage is estimated through template fitting to the various timestreams. We find the $T\rightarrow P$ leakage in the demodulated data is $<3.8\times 10^{-4}$ (95\% confidence) across the focal plane. The median knee frequencies of the linear demodulated data is 15.1~mHz, corresponding to a factor of 110 reduction in integration time at 1~mHz compared to pair-differencing but higher than would be expected from the $T \rightarrow P$ alone. The modulation efficiency, the ratio of white noise in the demodulated data to white noise of the data before demodulation, is found to be in good agreement with the theoretical prediction.


Lastly, we examine the demodulated linear timestreams to constrain possible sources of the $1/f$ noise in these data. We find the pixels closest to the forebaffle at the edge of the focal plane have significantly higher $1/f$ knees and that all knee frequencies increase for wind speeds above 4.25~m/s. The more central pixels have median knee frequency of $11.65\pm0.1$~mHz. This $1/f$ behavior further depends on PWV and the boresight angle of the observation. The PWV dependence, due to $T\rightarrow P$ leakages, accounts for about $32\%$ of the red noise in the CLASS 40~GHz data. We can expect this fraction to be higher for the higher observing frequencies where the atmospheric signal is brighter. A boresight dependent effect, likely due to mount vibrations, accounts for about 15\% of the $1/f$ noise. The is a remaining sources of $1/f$ noise that is likely associated with the instrument scanning motion because the stare data PSDs are found to have less $1/f$ noise than the scanning PSDs.

The $1/f$ behavior studied in this analysis presents the status of the data with no post-processing beyond the removal of an azimuth-synchronous fit. With the presence of a front-end VPM, the CLASS 40~GHz telescope has achieved a level of $1/f$ noise that, using the telescope scan strategy, brings the projected knee in multipole space well into the largest angular scales on the sky ($\ell \lesssim10$). This result is in agreement with the white noise levels observed in preliminary maps of the 40 GHz CLASS data. This would not be possible without the substantial reduction of $1/f$ noise due to the VPM. In future work, we will be expanding this analysis to include simultaneous multi-frequency observations to further constrain the sources of $1/f$ noise in the CLASS data. 


\section*{Acknowledgments}
\vskip4pt

We acknowledge the National Science Foundation Division of Astronomical Sciences for their support of CLASS under grant Nos. 0959349, 1429236, 1636634, 1654494, and 2034400. We thank Johns Hopkins University President R. Daniels and Dean J. Toscano for their steadfast support of CLASS. The CLASS project employs detector technology developed in collaboration between JHU and Goddard Space Flight Center under several previous and ongoing NASA grants. Detector development work at JHU was funded by NASA grant No. NNX14AB76A. K. Harrington was supported by NASA Space Technology Research Fellowship grant No. NX14AM49H and by the U.S. Department of Energy, Office of Science, under Award Number DE-SC0015799 for parts of this work. Zhilei Xu is supported by the Gordon and Betty Moore Foundation. R.D. and R.R. thanks CONICYT for grant BASAL CATA AFB-170002. R.R. acknowledges support from ANID-FONDECYT through grant 1181620. We thank scientists from NIST for their contributions to the detector and readout systems, including Johannes Hubmayr, Gene Hilton, and Carl Reintsema. We acknowledge scientific and engineering contributions from Max Abitbol, Fletcher Boone, David Carcamo, Saianeesh Haridas, Connor Henley, Lindsay Lowry, Isu Ravi, Gary Rhodes, Daniel Swartz, Bingjie Wang, Qinan Wang, Tiffany Wei, and Zi'ang Yan. We thank William Deysher, Mar\'ia Jos\'e Amaral, and Chantal Boisvert for logistical support. We acknowledge productive collaboration with Dean Carpenter and the JHU Physical Sciences Machine Shop team. We further acknowledge the very generous support of Jim and Heather Murren (JHU A\&S '88), Matthew Polk (JHU A\&S Physics BS '71), David Nicholson, and Michael Bloomberg (JHU Engineering '64). CLASS is located in the Parque Astron\'omico Atacama in northern Chile under the auspices of the Agencia Nacional de Investigaci\'on y Desarrollo (ANID).

\software{\texttt{IPython} \citep{ipython}, \texttt{numpy} \citep{numpy}, \texttt{scipy} \citep{scipy}, \texttt{matplotlib} \citep{matplotlib}, \texttt{PyEphem} \citep{pyephem}, 
\texttt{moby2} (\url{https://github.com/ACTCollaboration/moby2}),
}

\appendix
\section{\label{sec:single_dets} Sources of Intensity \lowercase{$1/f$} Noise}

\begin{figure*}
    \centering
    \begin{tabular}{cc}
    \includegraphics[width=0.42\textwidth]{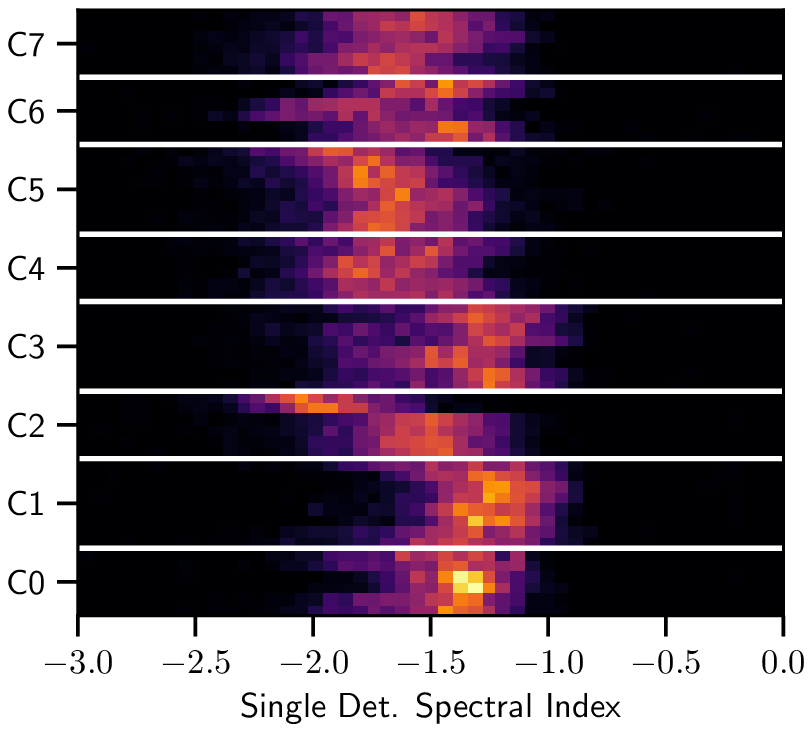} & 
    \includegraphics[width=0.48\textwidth]{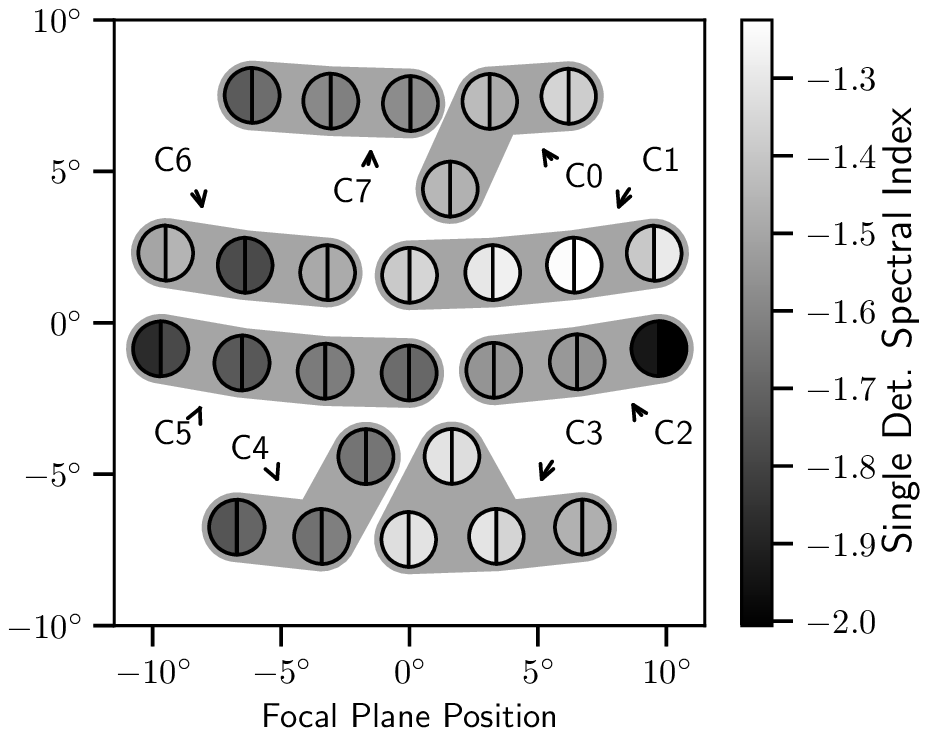}\\
    \end{tabular}
    \caption{\label{fig:single_det_alpha} (Left) Histograms of the \textit{single detector} spectral index distributions broken out per detector and grouped by readout column. (Right) The median single detector spectral indexes in the CLASS 40~GHz focal plane. Each half circle in this plot represents one of the detectors. The gray-scale half circles are shaded by the median single detector spectral index value. The shaded outer regions label the readout column assigned to each detector in the time-division multiplexing scheme. If atmospheric turbulence was the dominant source of $1/f$ noise we would expect the spectral index to be consistent across the focal plane. Larger scale atmospheric fluctuations would depend on airmass and elevation on the sky, meaning detectors with the same elevation offset in the focal plane would have the same spectral index. Although boresight rotations move the elevation offset to different elevations on the sky and this plot shows the median for all boresights, the observed pattern persists when the data are split by boresight. Instead, detectors with the same elevation offset but different readout column placement show different $1/f$ spectral indexes. This points to readout effects being a larger contributor to $1/f$ than the atmosphere for single detector data on the 2-hour timescales examined here. The stability of pair-differenced and pair-differenced demodulated data does not show a similar dependence on readout column.
    }
\end{figure*}

The single detector data and PSDs contain the atmospheric temperature information and any other intensity related signals. The atmosphere is normally the largest temperature source continuously observed by a ground-based CMB telescope. Any $T\rightarrow P$ leakage will convert atmospheric temperature into polarization, and the single detector data are one mechanism of calibrating the level of atmospheric signal observed by the CLASS 40~GHz telescope. This calibration is necessary because the CLASS 40~GHz beam, bandpass, and scan strategy are unique among operating Atacama telescopes. 

The observed atmospheric signal is comprised of a bulk airmass with a temperature that varies smoothly with a direction and time as well as small-scale turbulence modes. Atmospheric turbulence and the theoretical models describing it have been well characterized from the CLASS Atacama site \citep{Church1995,LayHalverson2000, Dunner2013,Errard2015}.

One of the signatures of atmospheric turbulence is the spectral index of the $f^{\alpha}$ power law in the data PSD. For Kolmogorov turbulence, this index ranges from $-8/3$ to $-11/3$ depending on the size scale of the turbulence modes \citep{LayHalverson2000}. This power law index and its variation with precipitable water vapor (PWV) have been observed from the Atacama by ACT \citep{Dunner2013} and from the South Pole by ACBAR \citep{Bussmann2005}. From Table \ref{tab:demod_stats}, the median spectral index observed at 40~GHz by the CLASS telescope is much shallower than the values predicted from turbulent motion in the atmosphere.

\cite{Errard2015} used \textsc{Polarbear} data to calibrate a model of turbulence for the Atacama and found the injection scale of the turbulent modes to be $\sim200-500$~m in size. Depending on the height where the turbulent modes are injected, this injection scale corresponds to an angular scale of $14^\circ - 0.3^\circ$ for the largest turbulent features. The 40~GHz CLASS beams have a full-width half max (FWHM) of $\sim1.6^\circ$ \citep{xu20}, meaning the $1/f$ power law observed by larger-aperture telescopes in the Atacama should not be expected for the CLASS 40~GHz telescope because the beams smooth over the majority of the turbulent modes in the atmosphere. 


Figure \ref{fig:single_det_alpha} shows plots of the spectral index distributions (left) and median spectral index (right) fit for each detector in the 40~GHz focal plane. The shaded regions in the right plot denote the assigned multiplexing column. An important feature in the right plot is that many of the closest pixel-pairs have significantly different median spectral indexes (e.g. the center of the bottom two rows). If the primary source of these long-timescale fluctuations was atmospheric, we would expect detectors next to each other to see the same atmosphere as the telescope scans at a constant elevation. The significant difference in spectral index between some nearest-neighbor pixels points to a source different than the atmosphere. 

A possible alternative source for the single-detector data long-timescale $1/f$ noise is the electronics in the focal plane and multiplexing readout system. Voltage fluctuations across the detector bias lines were observed to create correlated noise between detectors in the same readout column in dark detector testing for the 40~GHz CLASS focal plane. These voltage fluctuations would be expected to occur on all timescales. The nearest-neighbor pixel pairs with the most different spectral indexes are on multiplexing column boundaries and in-focal-plane-wiring, while the spectral indexes within columns are generally uniform. Since the atmospheric turbulent spectrum is expected to plateau at larger angular scales, and thus lower temporal frequencies, it appears the single detector $1/f$ is dominated by the fluctuations in the readout system. 


The column correlated noise is most easily removed from the data by pair-differencing, since every detector is on the same column as its pixel pair. For this reason, 40~GHz CLASS analysis, including the demodulated data in this paper, is done with pair-differenced data. However, the dominance of readout noise at long timescales will affect any measurements of intensity that requires pair-adding data. In particular, comparing the knee frequencies of single detector data to pair-differenced and/or demodulated data does not produce a reliable estimate of temperature to polarization leakage. A similar search of readout-induced $1/f$ was done for the demodulated data and no column-correlated effects could be found. This indicates that readout may be a dominant source of $1/f$ noise for single-detector data but not for pair-differenced demodulated data.


\bibliography{class,extra_citations,Planck_bib}

\begin{thebibliography}{}
\expandafter\ifx\csname natexlab\endcsname\relax\def\natexlab#1{#1}\fi
\providecommand{\url}[1]{\href{#1}{#1}}

\bibitem[{{Adachi} {et~al.}(2020){Adachi}, {Aguilar Fa{\'u}ndez}, {Arnold},
  {Baccigalupi}, {Barron}, {Beck}, {Bianchini}, {Chapman}, {Cheung}, {Chinone},
  {Crowley}, {Dobbs}, {El Bouhargani}, {Elleflot}, {Errard}, {Fabbian}, {Feng},
  {Fujino}, {Galitzki}, {Goeckner-Wald}, {Groh}, {Hall}, {Hasegawa}, {Hazumi},
  {Hirose}, {Jaffe}, {Jeong}, {Kaneko}, {Katayama}, {Keating}, {Kikuchi},
  {Kisner}, {Kusaka}, {Lee}, {Leon}, {Linder}, {Lowry}, {Matsuda}, {Matsumura},
  {Minami}, {Navaroli}, {Nishino}, {Pham}, {Poletti}, {Reichardt}, {Segawa},
  {Siritanasak}, {Tajima}, {Takakura}, {Takatori}, {Tanabe}, {Teply}, {Tsai},
  {Verg{\`e}s}, {Westbrook}, \& {Zhou}}]{adac20}
{Adachi}, S., {Aguilar Fa{\'u}ndez}, M.~A.~O., {Arnold}, K., {et~al.} 2020,
  arXiv e-prints, arXiv:2005.06168

\bibitem[{Appel {et~al.}(2019)Appel, Xu, Padilla, Harrington, Marquez, Ali,
  Bennett, Brewer, Bustos, Chan, Chuss, Cleary, Couto, Dahal, Denis, Dünner,
  Eimer, Essinger-Hileman, Fluxa, Gothe, Hilton, Hubmayr, Iuliano, Karakla,
  Marriage, Miller, N{\'{u}}{\~{n}}ez, Parker, Petroff, Reintsema, Rostem,
  Stevens, Valle, Wang, Watts, Wollack, \& Zeng}]{appe19}
Appel, J.~W., Xu, Z., Padilla, I.~L., {et~al.} 2019, \apj, 876, 126.
\newblock \url{https://doi.org/10.3847\%2F1538-4357\%2Fab1652}

\bibitem[{{Bennett} {et~al.}(2003){Bennett}, {Halpern}, {Hinshaw}, {Jarosik},
  {Kogut}, {Limon}, {Meyer}, {Page}, {Spergel}, {Tucker}, {Wollack}, {Wright},
  {Barnes}, {Greason}, {Hill}, {Komatsu}, {Nolta}, {Odegard}, {Peiris},
  {Verde}, \& {Weiland}}]{wmap_yr1_bennett}
{Bennett}, C.~L., {Halpern}, M., {Hinshaw}, G., {et~al.} 2003, \apjs, 148, 1

\bibitem[{{Bennett} {et~al.}(2013){Bennett}, {Larson}, {Weiland}, {Jarosik},
  {Hinshaw}, {Odegard}, {Smith}, {Hill}, {Gold}, {Halpern}, {Komatsu}, {Nolta},
  {Page}, {Spergel}, {Wollack}, {Dunkley}, {Kogut}, {Limon}, {Meyer}, {Tucker},
  \& {Wright}}]{bennett:2013}
{Bennett}, C.~L., {Larson}, D., {Weiland}, J.~L., {et~al.} 2013, \apjs, 208, 20

\bibitem[{{Bussmann} {et~al.}(2005){Bussmann}, {Holzapfel}, \&
  {Kuo}}]{Bussmann2005}
{Bussmann}, R.~S., {Holzapfel}, W.~L., \& {Kuo}, C.~L. 2005, \apj, 622, 1343

\bibitem[{{Bustos} {et~al.}(2014){Bustos}, {Rubio}, {Ot{\'a}rola}, \&
  {Nagar}}]{parque_atacama}
{Bustos}, R., {Rubio}, M., {Ot{\'a}rola}, A., \& {Nagar}, N. 2014, Publications
  of the Astronomical Society of the Pacific, 126, 1126

\bibitem[{Church(1995)}]{Church1995}
Church, S.~E. 1995, Monthly Notices of the Royal Astronomical Society, 272,
  551.
\newblock \url{https://doi.org/10.1093/mnras/272.3.551}

\bibitem[{{Chuss} {et~al.}(2012){Chuss}, {Wollack}, {Henry}, {Hui}, {Juarez},
  {Krejny}, {Moseley}, \& {Novak}}]{chus12}
{Chuss}, D.~T., {Wollack}, E.~J., {Henry}, R., {et~al.} 2012, \ao, 51, 197

\bibitem[{{Doroshkevich} {et~al.}(1978){Doroshkevich}, {Zel'dovich}, \&
  {Syunyaev}}]{doro78}
{Doroshkevich}, A.~G., {Zel'dovich}, Y.~B., \& {Syunyaev}, R.~A. 1978, \sovast,
  22, 523

\bibitem[{{D{\"u}nner} {et~al.}(2013){D{\"u}nner}, {Hasselfield}, {Marriage},
  {Sievers}, {Acquaviva}, {Addison}, {Ade}, {Aguirre}, {Amiri}, {Appel},
  {Barrientos}, {Battistelli}, {Bond}, {Brown}, {Burger}, {Calabrese},
  {Chervenak}, {Das}, {Devlin}, {Dicker}, {Bertrand Doriese}, {Dunkley},
  {Essinger-Hileman}, {Fisher}, {Gralla}, {Fowler}, {Hajian}, {Halpern},
  {Hern{\'a}ndez-Monteagudo}, {Hilton}, {Hilton}, {Hincks}, {Hlozek},
  {Huffenberger}, {Hughes}, {Hughes}, {Infante}, {Irwin}, {Baptiste Juin},
  {Kaul}, {Klein}, {Kosowsky}, {Lau}, {Limon}, {Lin}, {Louis}, {Lupton},
  {Marsden}, {Martocci}, {Mauskopf}, {Menanteau}, {Moodley}, {Moseley},
  {Netterfield}, {Niemack}, {Nolta}, {Page}, {Parker}, {Partridge}, {Quintana},
  {Reid}, {Sehgal}, {Sherwin}, {Spergel}, {Staggs}, {Swetz}, {Switzer},
  {Thornton}, {Trac}, {Tucker}, {Warne}, {Wilson}, {Wollack}, \&
  {Zhao}}]{Dunner2013}
{D{\"u}nner}, R., {Hasselfield}, M., {Marriage}, T.~A., {et~al.} 2013, \apj,
  762, 10

\bibitem[{{Eimer} {et~al.}(2012){Eimer}, {Bennett}, {Chuss}, {Marriage},
  {Wollack}, \& {Zeng}}]{eime12}
{Eimer}, J.~R., {Bennett}, C.~L., {Chuss}, D.~T., {et~al.} 2012, in Society of
  Photo-Optical Instrumentation Engineers (SPIE) Conference Series, Vol. 8452,
  Society of Photo-Optical Instrumentation Engineers (SPIE) Conference Series

\bibitem[{{Errard} {et~al.}(2015){Errard}, {Ade}, {Akiba}, {Arnold}, {Atlas},
  {Baccigalupi}, {Barron}, {Boettger}, {Borrill}, {Chapman}, {Chinone},
  {Cukierman}, {Delabrouille}, {Dobbs}, {Ducout}, {Elleflot}, {Fabbian},
  {Feng}, {Feeney}, {Gilbert}, {Goeckner-Wald}, {Halverson}, {Hasegawa},
  {Hattori}, {Hazumi}, {Hill}, {Holzapfel}, {Hori}, {Inoue}, {Jaehnig},
  {Jaffe}, {Jeong}, {Katayama}, {Kaufman}, {Keating}, {Kermish}, {Keskitalo},
  {Kisner}, {Le Jeune}, {Lee}, {Leitch}, {Leon}, {Linder}, {Matsuda},
  {Matsumura}, {Miller}, {Myers}, {Navaroli}, {Nishino}, {Okamura}, {Paar},
  {Peloton}, {Poletti}, {Puglisi}, {Rebeiz}, {Reichardt}, {Richards}, {Ross},
  {Rotermund}, {Schenck}, {Sherwin}, {Siritanasak}, {Smecher}, {Stebor},
  {Steinbach}, {Stompor}, {Suzuki}, {Tajima}, {Takakura}, {Tikhomirov},
  {Tomaru}, {Whitehorn}, {Wilson}, {Yadav}, \& {Zahn}}]{Errard2015}
{Errard}, J., {Ade}, P.~A.~R., {Akiba}, Y., {et~al.} 2015, \apj, 809, 63

\bibitem[{{Essinger-Hileman} {et~al.}(2014){Essinger-Hileman}, {Ali}, {Amiri},
  {Appel}, {Araujo}, {Bennett}, {Boone}, {Chan}, {Cho}, {Chuss}, {Colazo},
  {Crowe}, {Denis}, {D{\"u}nner}, {Eimer}, {Gothe}, {Halpern}, {Harrington},
  {Hilton}, {Hinshaw}, {Huang}, {Irwin}, {Jones}, {Karakla}, {Kogut}, {Larson},
  {Limon}, {Lowry}, {Marriage}, {Mehrle}, {Miller}, {Miller}, {Moseley},
  {Novak}, {Reintsema}, {Rostem}, {Stevenson}, {Towner}, {U-Yen}, {Wagner},
  {Wollack}, {Xu}, \& {Zeng}}]{essi14}
{Essinger-Hileman}, T., {Ali}, A., {Amiri}, M., {et~al.} 2014, in SPIE, Vol.
  915354, Millimeter, Submillimeter, and Far-Infrared Detectors and
  Instrumentation for Astronomy VII

\bibitem[{{Essinger-Hileman} {et~al.}(2016){Essinger-Hileman}, {Kusaka},
  {Appel}, {Choi}, {Crowley}, {Ho}, {Jarosik}, {Page}, {Parker}, {Raghunathan},
  {Simon}, {Staggs}, \& {Visnjic}}]{essinger2016}
{Essinger-Hileman}, T., {Kusaka}, A., {Appel}, J.~W., {et~al.} 2016, Review of
  Scientific Instruments, 87, 094503

\bibitem[{{Hanany} \& {Rosenkranz}(2003)}]{Hanany2003}
{Hanany}, S., \& {Rosenkranz}, P. 2003, \nar, 47, 1159

\bibitem[{{Harrington} {et~al.}(2016){Harrington}, {Marriage}, {Ali}, {Appel},
  {Bennett}, {Boone}, {Brewer}, {Chan}, {Chuss}, {Colazo}, {Dahal}, {Denis},
  {D{\"u}nner}, {Eimer}, {Essinger-Hileman}, {Fluxa}, {Halpern}, {Hilton},
  {Hinshaw}, {Hubmayr}, {Iuliano}, {Karakla}, {McMahon}, {Miller}, {Moseley},
  {Palma}, {Parker}, {Petroff}, {Pradenas}, {Rostem}, {Sagliocca}, {Valle},
  {Watts}, {Wollack}, {Xu}, \& {Zeng}}]{harr16}
{Harrington}, K., {Marriage}, T., {Ali}, A., {et~al.} 2016, in \procspie, Vol.
  9914, Millimeter, Submillimeter, and Far-Infrared Detectors and
  Instrumentation for Astronomy VIII, 99141K

\bibitem[{{Harrington} {et~al.}(2018){Harrington}, {Eimer}, {Chuss}, {Petroff},
  {Cleary}, {DeGeorge}, {Grunberg}, {Ali}, {Appel}, {Bennett}, {Brewer},
  {Bustos}, {Chan}, {Couto}, {Dahal}, {Denis}, {D{\"u}nner},
  {Essinger-Hileman}, {Fluxa}, {Halpern}, {Hilton}, {Hinshaw}, {Hubmayr},
  {Iuliano}, {Karakla}, {Marriage}, {McMahon}, {Miller}, {Nu{\~n}ez},
  {Padilla}, {Palma}, {Parker}, {Pradenas Marquez}, {Reeves}, {Reintsema},
  {Rostem}, {Augusto Nunes Valle}, {Van Engelhoven}, {Wang}, {Wang}, {Watts},
  {Weiland}, {Wollack}, {Xu}, {Yan}, \& {Zeng}}]{harr18}
{Harrington}, K., {Eimer}, J., {Chuss}, D.~T., {et~al.} 2018, in Society of
  Photo-Optical Instrumentation Engineers (SPIE) Conference Series, Vol. 10708,
  Society of Photo-Optical Instrumentation Engineers (SPIE) Conference Series,
  107082M

\bibitem[{{Harrington}(2018)}]{Harrington_Thesis_2018}
{Harrington}, K.~M. 2018, PhD thesis, Johns Hopkins University

\bibitem[{Hecht(2012)}]{hecht2012optics}
Hecht, E. 2012, Optics (Pearson)

\bibitem[{{Heinrich} \& {Hu}(2018)}]{heinrich}
{Heinrich}, C., \& {Hu}, W. 2018, \prd, 98, 063514

\bibitem[{{Hinshaw} {et~al.}(2013){Hinshaw}, {Larson}, {Komatsu}, {Spergel},
  {Bennett}, {Dunkley}, {Nolta}, {Halpern}, {Hill}, {Odegard}, {Page}, {Smith},
  {Weiland}, {Gold}, {Jarosik}, {Kogut}, {Limon}, {Meyer}, {Tucker}, {Wollack},
  \& {Wright}}]{Hinshaw2013}
{Hinshaw}, G., {Larson}, D., {Komatsu}, E., {et~al.} 2013, \apjs, 208, 19

\bibitem[{{Hu} \& {Holder}(2003)}]{hu}
{Hu}, W., \& {Holder}, G.~P. 2003, \prd, 68, 023001

\bibitem[{Hunter(2007)}]{matplotlib}
Hunter, J.~D. 2007, CSE, 9, 90

\bibitem[{{Kamionkowski} {et~al.}(1997){Kamionkowski}, {Kosowsky}, \&
  {Stebbins}}]{kami97}
{Kamionkowski}, M., {Kosowsky}, A., \& {Stebbins}, A. 1997, \prd, 55, 7368

\bibitem[{{Keck Array} {et~al.}(2018){Keck Array}, {BICEP2 Collaborations},
  {:}, {Ade}, {Ahmed}, {Aikin}, {Alexander}, {Barkats}, {Benton}, {Bischoff},
  {Bock}, {Bowens-Rubin}, {Brevik}, {Buder}, {Bullock}, {Buza}, {Connors},
  {Cornelison}, {Crill}, {Crumrine}, {Dierickx}, {Duband}, {Dvorkin},
  {Filippini}, {Fliescher}, {Grayson}, {Hall}, {Halpern}, {Harrison},
  {Hildebrandt}, {Hilton}, {Hui}, {Irwin}, {Kang}, {Karkare}, {Karpel},
  {Kaufman}, {Keating}, {Kefeli}, {Kernasovskiy}, {Kovac}, {Kuo}, {Larsen},
  {Lau}, {Leitch}, {Lueker}, {Megerian}, {Moncelsi}, {Namikawa}, {Netterfield},
  {Nguyen}, {O'Brient}, {Ogburn}, {Palladino}, {Pryke}, {Racine}, {Richter},
  {Schillaci}, {Schwarz}, {Sheehy}, {Soliman}, {St.~Germaine}, {Staniszewski},
  {Steinbach}, {Sudiwala}, {Teply}, {Thompson}, {Tolan}, {Tucker}, {Turner},
  {Umilta}, {Vieregg}, {Wandui}, {Weber}, {Wiebe}, {Willmert}, {Wong}, {Wu},
  {Yang}, {Yoon}, \& {Zhang}}]{bicep2_2018_2015}
{Keck Array}, {BICEP2 Collaborations}, {:}, {et~al.} 2018, ArXiv e-prints,
  arXiv:1810.05216

\bibitem[{{Keih{\"a}nen} {et~al.}(2010){Keih{\"a}nen}, {Keskitalo},
  {Kurki-Suonio}, {Poutanen}, \& {Sirvi{\"o}}}]{madam2010}
{Keih{\"a}nen}, E., {Keskitalo}, R., {Kurki-Suonio}, H., {Poutanen}, T., \&
  {Sirvi{\"o}}, A.~S. 2010, \aap, 510, A57

\bibitem[{{Kusaka} {et~al.}(2014){Kusaka}, {Essinger-Hileman}, {Appel},
  {Gallardo}, {Irwin}, {Jarosik}, {Nolta}, {Page}, {Parker}, {Raghunathan},
  {Sievers}, {Simon}, {Staggs}, \& {Visnjic}}]{abs_hwp}
{Kusaka}, A., {Essinger-Hileman}, T., {Appel}, J.~W., {et~al.} 2014, Review of
  Scientific Instruments, 85, 024501

\bibitem[{{Kusaka} {et~al.}(2018){Kusaka}, {Appel}, {Essinger-Hileman},
  {Beall}, {Campusano}, {Cho}, {Choi}, {Crowley}, {Fowler}, {Gallardo},
  {Hasselfield}, {Hilton}, {Ho}, {Irwin}, {Jarosik}, {Niemack}, {Nixon},
  {\~{}Nolta}, {Page}, {Palma}, {Parker}, {Raghunathan}, {Reintsema},
  {Sievers}, {Simon}, {Staggs}, {Visnjic}, \& {Yoon}}]{abs_final}
{Kusaka}, A., {Appel}, J., {Essinger-Hileman}, T., {et~al.} 2018, \jcap, 9, 005

\bibitem[{{Lay} \& {Halverson}(2000)}]{LayHalverson2000}
{Lay}, O.~P., \& {Halverson}, N.~W. 2000, \apj, 543, 787

\bibitem[{{Louis} {et~al.}(2017){Louis}, {Grace}, {Hasselfield}, {Lungu},
  {Maurin}, {Addison}, {Ade}, {Aiola}, {Allison}, {Amiri}, {Angile},
  {Battaglia}, {Beall}, {de Bernardis}, {Bond}, {Britton}, {Calabrese}, {Cho},
  {Choi}, {Coughlin}, {Crichton}, {Crowley}, {Datta}, {Devlin}, {Dicker},
  {Dunkley}, {D{\"u}nner}, {Ferraro}, {Fox}, {Gallardo}, {Gralla}, {Halpern},
  {Henderson}, {Hill}, {Hilton}, {Hilton}, {Hincks}, {Hlozek}, {Ho}, {Huang},
  {Hubmayr}, {Huffenberger}, {Hughes}, {Infante}, {Irwin}, {Muya Kasanda},
  {Klein}, {Koopman}, {Kosowsky}, {Li}, {Madhavacheril}, {Marriage}, {McMahon},
  {Menanteau}, {Moodley}, {Munson}, {Naess}, {Nati}, {Newburgh}, {Nibarger},
  {Niemack}, {Nolta}, {Nu{\~n}ez}, {Page}, {Pappas}, {Partridge}, {Rojas},
  {Schaan}, {Schmitt}, {Sehgal}, {Sherwin}, {Sievers}, {Simon}, {Spergel},
  {Staggs}, {Switzer}, {Thornton}, {Trac}, {Treu}, {Tucker}, {Van Engelen},
  {Ward}, \& {Wollack}}]{loui17}
{Louis}, T., {Grace}, E., {Hasselfield}, M., {et~al.} 2017, Journal of
  Cosmology and Astro-Particle Physics, 2017, 031

\bibitem[{{Miller} {et~al.}(2016){Miller}, {Chuss}, {Marriage}, {Wollack},
  {Appel}, {Bennett}, {Eimer}, {Essinger-Hileman}, {Fixsen}, {Harrington},
  {Moseley}, {Rostem}, {Switzer}, \& {Watts}}]{mill15}
{Miller}, N.~J., {Chuss}, D.~T., {Marriage}, T.~A., {et~al.} 2016, \apj, 818,
  151

\bibitem[{{Obied} {et~al.}(2018){Obied}, {Dvorkin}, {Heinrich}, {Hu}, \&
  {Miranda}}]{obied}
{Obied}, G., {Dvorkin}, C., {Heinrich}, C., {Hu}, W., \& {Miranda}, V. 2018,
  \prd, 98, 043518

\bibitem[{{Padilla} {et~al.}(2020){Padilla}, {Eimer}, {Li}, {Addison}, {Ali},
  {Appel}, {Bennett}, {Bustos}, {Brewer}, {Chan}, {Chuss}, {Cleary}, {Couto},
  {Dahal}, {Denis}, {D{\"u}nner}, {Essinger-Hileman}, {Flux{\'a}}, {Gothe},
  {Haridas}, {Harrington}, {Iuliano}, {Karakla}, {Marriage}, {Miller},
  {N{\'u}{\~n}ez}, {Parker}, {Petroff}, {Reeves}, {Rostem}, {Stevens}, {Nunes
  Valle}, {Watts}, {Weiland}, {Wollack}, \& {Xu}}]{padi20}
{Padilla}, I.~L., {Eimer}, J.~R., {Li}, Y., {et~al.} 2020, \apj, 889, 105

\bibitem[{{Page} {et~al.}(2007){Page}, {Hinshaw}, {Komatsu}, {Nolta},
  {Spergel}, {Bennett}, {Barnes}, {Bean}, {Dor{\'e}}, {Dunkley}, {Halpern},
  {Hill}, {Jarosik}, {Kogut}, {Limon}, {Meyer}, {Odegard}, {Peiris}, {Tucker},
  {Verde}, {Weiland}, {Wollack}, \& {Wright}}]{Page2007}
{Page}, L., {Hinshaw}, G., {Komatsu}, E., {et~al.} 2007, \apjs, 170, 335

\bibitem[{{Papadakis} \& {Lawrence}(1993)}]{Papadakis1993}
{Papadakis}, I.~E., \& {Lawrence}, A. 1993, \mnras, 261, 612

\bibitem[{{Peebles} \& {Yu}(1970)}]{peeb70}
{Peebles}, P.~J.~E., \& {Yu}, J.~T. 1970, \apj, 162, 815

\bibitem[{P\'erez \& Granger(2007)}]{ipython}
P\'erez, F., \& Granger, B.~E. 2007, CSE, 9, 21.
\newblock \url{http://ipython.org}

\bibitem[{{Petroff} {et~al.}(2020){Petroff}, {Eimer}, {Harrington}, {Ali},
  {Appel}, {Bennett}, {Brewer}, {Bustos}, {Chan}, {Chuss}, {Cleary}, {Couto},
  {Dahal}, {D{\"u}nner}, {Essinger-Hileman}, {Rojas}, {Gothe}, {Iuliano},
  {Marriage}, {Miller}, {N{\'u}{\~n}ez}, {Padilla}, {Parker}, {Reeves},
  {Rostem}, {Nunes Valle}, {Watts}, {Weiland}, {Wollack}, \& {Xu}}]{petr20}
{Petroff}, M.~A., {Eimer}, J.~R., {Harrington}, K., {et~al.} 2020, \apj, 889,
  120

\bibitem[{{\sorthelp{Planck Collaboration 2018A}}{Planck Collaboration
  I}(2019)}]{planck2016-l01}
{\sorthelp{Planck Collaboration 2018A}}{Planck Collaboration I}. 2019, \aap, in
  press, arXiv:1807.06205

\bibitem[{{\sorthelp{Planck Collaboration 2018F}}{Planck Collaboration
  VI}(2019)}]{planck2016-l06}
{\sorthelp{Planck Collaboration 2018F}}{Planck Collaboration VI}. 2019, \aap,
  in press, arXiv:1807.06209

\bibitem[{{Poletti} {et~al.}(2017){Poletti}, {Fabbian}, {Le Jeune}, {Peloton},
  {Arnold}, {Baccigalupi}, {Barron}, {Beckman}, {Borrill}, {Chapman},
  {Chinone}, {Cukierman}, {Ducout}, {Elleflot}, {Errard}, {Feeney},
  {Goeckner-Wald}, {Groh}, {Hall}, {Hasegawa}, {Hazumi}, {Hill}, {Howe},
  {Inoue}, {Jaffe}, {Jeong}, {Katayama}, {Keating}, {Keskitalo}, {Kisner},
  {Kusaka}, {Lee}, {Leon}, {Linder}, {Lowry}, {Matsuda}, {Navaroli}, {Paar},
  {Puglisi}, {Reichardt}, {Ross}, {Siritanasak}, {Stebor}, {Steinbach},
  {Stompor}, {Suzuki}, {Tajima}, {Teply}, \& {Whitehorn}}]{poletti2016}
{Poletti}, D., {Fabbian}, G., {Le Jeune}, M., {et~al.} 2017, \aap, 600, A60

\bibitem[{{QUIET Collaboration} {et~al.}(2011){QUIET Collaboration},
  {Bischoff}, {Brizius}, {Buder}, {Chinone}, {Cleary}, {Dumoulin}, {Kusaka},
  {Monsalve}, {N{\ae}ss}, {Newburgh}, {Reeves}, {Smith}, {Wehus}, {Zuntz},
  {Zwart}, {Bronfman}, {Bustos}, {Church}, {Dickinson}, {Eriksen}, {Ferreira},
  {Gaier}, {Gundersen}, {Hasegawa}, {Hazumi}, {Huffenberger}, {Jones},
  {Kangaslahti}, {Kapner}, {Lawrence}, {Limon}, {May}, {McMahon}, {Miller},
  {Nguyen}, {Nixon}, {Pearson}, {Piccirillo}, {Radford}, {Readhead},
  {Richards}, {Samtleben}, {Seiffert}, {Shepherd}, {Staggs}, {Tajima},
  {Thompson}, {Vand erlinde}, {Williamson}, \& {Winstein}}]{QUIET_results2011}
{QUIET Collaboration}, {Bischoff}, C., {Brizius}, A., {et~al.} 2011, \apj, 741,
  111

\bibitem[{{Rhodes}(2011)}]{pyephem}
{Rhodes}, B.~C. 2011, {PyEphem: Astronomical Ephemeris for Python},
  Astrophysics Source Code Library, vv3.7.6.0, , , ascl:1112.014

\bibitem[{{Seljak} \& {Zaldarriaga}(1997)}]{Sel_Zal1997}
{Seljak}, U., \& {Zaldarriaga}, M. 1997, \prl, 78, 2054

\bibitem[{{Spergel} {et~al.}(2003){Spergel}, {Verde}, {Peiris}, {Komatsu},
  {Nolta}, {Bennett}, {Halpern}, {Hinshaw}, {Jarosik}, {Kogut}, {Limon},
  {Meyer}, {Page}, {Tucker}, {Weiland}, {Wollack}, \&
  {Wright}}]{wmap_yr1_spergel}
{Spergel}, D.~N., {Verde}, L., {Peiris}, H.~V., {et~al.} 2003, \apjs, 148, 175

\bibitem[{{Takakura} {et~al.}(2017){Takakura}, {Aguilar}, {Akiba}, {Arnold},
  {Baccigalupi}, {Barron}, {Beckman}, {Boettger}, {Borrill}, {Chapman},
  {Chinone}, {Cukierman}, {Ducout}, {Elleflot}, {Errard}, {Fabbian}, {Fujino},
  {Galitzki}, {Goeckner-Wald}, {Halverson}, {Hasegawa}, {Hattori}, {Hazumi},
  {Hill}, {Howe}, {Inoue}, {Jaffe}, {Jeong}, {Kaneko}, {Katayama}, {Keating},
  {Keskitalo}, {Kisner}, {Krachmalnicoff}, {Kusaka}, {Lee}, {Leon}, {Lowry},
  {Matsuda}, {Matsumura}, {Navaroli}, {Nishino}, {Paar}, {Peloton}, {Poletti},
  {Puglisi}, {Reichardt}, {Ross}, {Siritanasak}, {Suzuki}, {Tajima},
  {Takatori}, \& {Teply}}]{Takakura2017}
{Takakura}, S., {Aguilar}, M., {Akiba}, Y., {et~al.} 2017, \jcap, 2017, 008

\bibitem[{{Takakura} {et~al.}(2019){Takakura}, {Aguilar-Fa{\'u}ndez}, {Akiba},
  {Arnold}, {Baccigalupi}, {Barron}, {Beck}, {Bianchini}, {Boettger},
  {Borrill}, {Cheung}, {Chinone}, {Elleflot}, {Errard}, {Fabbian}, {Feng},
  {Goeckner-Wald}, {Hamada}, {Hasegawa}, {Hazumi}, {Howe}, {Kaneko},
  {Katayama}, {Keating}, {Keskitalo}, {Kisner}, {Krachmalnicoff}, {Kusaka},
  {Lee}, {Lowry}, {Matsuda}, {May}, {Minami}, {Navaroli}, {Nishino},
  {Piccirillo}, {Poletti}, {Puglisi}, {Reichardt}, {Segawa}, {Silva-Feaver},
  {Siritanasak}, {Suzuki}, {Tajima}, {Takatori}, {Tanabe}, {Teply}, \&
  {Tsai}}]{Takakura2019}
{Takakura}, S., {Aguilar-Fa{\'u}ndez}, M.~A.~O., {Akiba}, Y., {et~al.} 2019,
  \apj, 870, 102

\bibitem[{van~der Walt {et~al.}(2011)van~der Walt, Colbert, \&
  Varoquaux}]{numpy}
van~der Walt, S., Colbert, S.~C., \& Varoquaux, G. 2011, CSE, 13, 22.
\newblock \url{https://doi.org/10.1109\%2Fmcse.2011.37}

\bibitem[{{Virtanen} {et~al.}(2019){Virtanen}, {Gommers}, {Oliphant},
  {Haberland}, {Reddy}, {Cournapeau}, {Burovski}, {Peterson}, {Weckesser},
  {Bright}, {van der Walt}, {Brett}, {Wilson}, {Jarrod Millman}, {Mayorov},
  {Nelson}, {Jones}, {Kern}, {Larson}, {Carey}, {Polat}, {Feng}, {Moore}, {Vand
  erPlas}, {Laxalde}, {Perktold}, {Cimrman}, {Henriksen}, {Quintero}, {Harris},
  {Archibald}, {Ribeiro}, {Pedregosa}, {van Mulbregt}, \& {SciPy 1.0
  Contributors}}]{scipy}
{Virtanen}, P., {Gommers}, R., {Oliphant}, T.~E., {et~al.} 2019, arXiv
  e-prints, arXiv:1907.10121

\bibitem[{{Watts} {et~al.}(2020){Watts}, {Addison}, {Bennett}, \& {Weiland
  }}]{watts2020}
{Watts}, D.~J., {Addison}, G.~E., {Bennett}, C.~L., \& {Weiland }, J.~L. 2020,
  \apj, 889, 130

\bibitem[{{Xu} {et~al.}(2020){Xu}, {Brewer}, {Rojas}, {Li}, {Osumi},
  {Pradenas}, {Ali}, {Appel}, {Bennett}, {Bustos}, {Chan}, {Chuss}, {Cleary},
  {Couto}, {Dahal}, {Datta}, {Denis}, {D{\"u}nner}, {Eimer},
  {Essinger-Hileman}, {Gothe}, {Harrington}, {Iuliano}, {Karakla}, {Marriage},
  {Miller}, {N{\'u}{\~n}ez}, {Padilla}, {Parker}, {Petroff}, {Reeves},
  {Rostem}, {Nunes Valle}, {Watts}, {Weiland }, {Wollack}, \& {CLASS
  Collaboration}}]{xu20}
{Xu}, Z., {Brewer}, M.~K., {Rojas}, P.~F., {et~al.} 2020, \apj, 891, 134

\end{thebibliography}
\end{document}